\documentclass{SciPost}

\hypersetup{
    colorlinks,
    linkcolor={red!50!black},
    citecolor={blue!50!black},
    urlcolor={blue!80!black}
}

\usepackage[bitstream-charter]{mathdesign}
\DeclareSymbolFont{usualmathcal}{OMS}{cmsy}{m}{n}
\DeclareSymbolFontAlphabet{\mathcal}{usualmathcal}

\fancypagestyle{SPstyle}{
\fancyhf{}
\lhead{\colorbox{scipostblue}{\bf \color{white} ~SciPost Physics }}
\rhead{{\bf \color{scipostdeepblue} ~Submission }}

\fancyfoot[C]{\textbf{\thepage}}
}

\usepackage{siunitx}

\usepackage{tikz}
\usetikzlibrary{calc} 
\usetikzlibrary{patterns} 
\usetikzlibrary{decorations.markings,arrows.meta} 

\def\centerarc[#1](#2)(#3:#4:#5){ \draw[#1] ($(#2)+({#5*cos(#3)},{#5*sin(#3)})$) arc (#3:#4:#5); }

\tikzset{->-/.style={decoration={
			markings,
			mark=at position #1 with {\arrow{latex[scale=1]}}}, postaction={decorate}}, ->-/.default=.5}

\def\hexagonsize{3.5} 
\pgfdeclarepatternformonly
  {hexagons}
  {\pgfpointorigin}
  {\pgfpoint{3*\hexagonsize}{0.866025*2*\hexagonsize}}
  {\pgfpoint{3*\hexagonsize}{0.866025*2*\hexagonsize}}
  {
   \pgfsetlinewidth{0.4pt}
   \pgftransformshift{\pgfpoint{0mm}{0.866025*\hexagonsize}}
   \pgfpathmoveto{\pgfpoint{0mm}{0mm}}
   \pgfpathlineto{\pgfpoint{0.5*\hexagonsize}{0mm}}
   \pgfpathlineto{\pgfpoint{\hexagonsize}{-0.866025*\hexagonsize}}
   \pgfpathlineto{\pgfpoint{2*\hexagonsize}{-0.866025*\hexagonsize}}
   \pgfpathlineto{\pgfpoint{2.5*\hexagonsize}{0mm}}
   \pgfpathlineto{\pgfpoint{3*\hexagonsize+0.2mm}{0mm}}
   \pgfpathmoveto{\pgfpoint{0.5*\hexagonsize}{0mm}}
   \pgfpathlineto{\pgfpoint{\hexagonsize}{0.866025*\hexagonsize}}
   \pgfpathlineto{\pgfpoint{2*\hexagonsize}{0.866025*\hexagonsize}}
   \pgfpathlineto{\pgfpoint{2.5*\hexagonsize}{0mm}}
   \pgfusepath{stroke}
  }

\begin{document}

\pagestyle{SPstyle}

\begin{center}{\Large \textbf{\color{scipostdeepblue}{
Sagnac and Mashhoon effects in graphene \\
}}}\end{center}

\begin{center}\textbf{
Yuri V. Shtanov\textsuperscript{1$\star$},
Taras-Hryhorii O. Pokalchuk\textsuperscript{2$\dagger$} and
Sergei G. Sharapov\textsuperscript{1,2$\ddagger$}
}\end{center}

\begin{center}
{\bf 1} Bogolyubov Institute for Theoretical Physics, 14-b Metrologichna st., Kyiv 03143, Ukraine
\\
{\bf 2} Kyiv Academic University, 36 Vernadsky blvd., Kyiv 03142, Ukraine
\\[\baselineskip]
$\star$ \href{mailto:shtanov@bitp.kyiv.ua}{\small shtanov@bitp.kyiv.ua}\,,\quad
$\dagger$ \href{mailto:taras.pokalchuk@gmail.com}{\small taras.pokalchuk@gmail.com}\,, \quad $\ddagger$ \href{mailto:sharapov@bitp.kyiv.ua}{\small sharapov@bitp.kyiv.ua}
\end{center}

\section*{\color{scipostdeepblue}{Abstract}}
\textbf{\boldmath{%
We investigate the Sagnac and Mashhoon effects in graphene, taking into account both the pseudospin and intrinsic spin of electrons, within a simplified model of a rotating nanotube or infinitesimally narrow ring. Based on considerations of the relativistic phase of the wave function and employing the effective Larmor theorem, we demonstrate that the Sagnac fringe shift retains a form analogous to that for free electrons, governed by the electron's vacuum mass. In the case of a narrow ring, an additional $\pi$-phase shift arises due to the Berry phase associated with the honeycomb graphene lattice. The Mashhoon fringe shift retains its conventional form, with its dependence on the Fermi velocity.
}}

\vspace{\baselineskip}



\vspace{10pt}
\noindent\rule{\textwidth}{1pt}
\tableofcontents
\noindent\rule{\textwidth}{1pt}
\vspace{10pt}


\section{Introduction}

Matter-wave interferometry serves as a powerful method for exploring quantum phenomena and their practical applications. A particularly clear manifestation of quantum interference in solid-state systems is the observation of periodic conductance oscillations in ring-shaped structures subjected to a magnetic field (see Ref.~\cite{Ihn2010book}). These oscillations arise from the Aharonov--Bohm effect, which reflects the phase difference accumulated by electron wavefunctions traveling along two distinct paths enclosing a magnetic flux. This phase difference is directly proportional to the magnetic flux enclosed by the paths and perpendicular to the plane of the ring, normalized by the magnetic flux quantum $\Phi_0 = 2\pi\hbar c/e$, where $\hbar$ is the reduced Planck constant, $-e <0$ is the electron charge, and $c$ is the speed of light in vacuum. The Aharonov--Bohm effect has been extensively studied in mesoscopic rings fabricated from metallic films and semiconductor heterostructures, contributing significantly to the development of mesoscopic physics.

Over the past two decades, graphene has emerged as an exceptional platform for studying the Aharonov--Bohm effect and other quantum interference phenomena, thanks to its long phase coherence length, on the order of several microns at temperatures below \SI{4}{K} \cite{Miao2007Science} (see also the reviews in Refs.~\cite{Schelter2012SSC, Chakraborti.review}). The first experimental observation of the Aharonov--Bohm effect in a two-terminal, gated ring structure fabricated from exfoliated single-layer graphene was reported in Ref.~\cite{Russo2008PRB}. Subsequent studies explored four-terminal resistance in similarly structured rings with side and back gates, revealing high visibility of Aharonov--Bohm oscillations, up to 10\% in amplitude \cite{Huefner2009PSS, Huefner2010NJP}. The Aharonov--Bohm effect was also investigated in graphene rings incorporating a p--n--p junction, where no significant change in the oscillation period or amplitude was observed in this dipolar regime \cite{Smirnov2012APL}.

A notable advancement came with the realization of an electron interferometer defined entirely by electrostatic gating in encapsulated bilayer graphene, which exhibited a phase coherence length exceeding that of etched devices \cite{Ensslin2022NanoLett}. More recently, Aharonov--Bohm oscillations were observed at \SI{4}{K} in graphene rings fabricated from chemical vapor deposition-grown graphene, marking a step forward in operational temperature compared to earlier exfoliated devices \cite{Tang2023IEEEOpen}. Finally, Aharonov--Bohm oscillations have been demonstrated in magic-angle twisted bilayer graphene for both dispersive and flat-band electrons \cite{Iwakiri2023NatCom}. The same moir{\'e} device also enabled observation of the Little–Parks effect within the superconducting phase, evidenced by oscillations in magnetoresistance and critical current, thereby confirming charge-$2e$ pairing.

It is worth noting that carbon nanotubes were studied well before the discovery of graphene \cite{Saito1998book}. Single-walled carbon nanotubes have long served as a theoretical benchmark for modeling the Aharonov--Bohm effect, due to their effectively one-dimensional nature, which allows the radial motion of electrons to be neglected \cite{Ajiki1993JPSJ, Saito1998book, Ando2005JPSJ}. The Aharonov--Bohm effect was experimentally observed in a suspended chiral single-walled carbon nanotubes by measuring its conductance under a magnetic field applied along the tube’s axis \cite{Cao2004PRL}. Aharonov--Bohm conductance oscillations were also reported in ballistic multi-walled carbon nanotubes \cite{Lassagne2007PRL}. While carbon nanotube experiments are highly sophisticated and have provided key insights into quantum interference, graphene offers a more versatile and tunable platform for exploring such phenomena.

While the Aharonov--Bohm effect arises from a static magnetic field enclosed by and perpendicular to the electron paths, interference patterns can also result from the Sagnac effect. It refers to the phenomenon in which a phase shift occurs between two coherent beams propagating in opposite directions within an interferometer that is rotating as a whole (see Refs.~\cite{Post1967RMP, Stedman1997RPP, Malykin2000UFN, Pascoli2017} for reviews).

Although originally discovered for light waves \cite{Sagnac1913a, Sagnac1913b}, the Sagnac effect is a general interference phenomenon that applies to matter waves of any kind. It has been experimentally demonstrated with a wide range of quantum particles, including superconducting Cooper pairs \cite{Zimmerman1965PRL}, neutrons \cite{Werner1979PRL}, and neutral atoms such as $^{40}$Ca \cite{Riehle1991PRL}. More recently, the effect was observed in a Cesium atom interferometer \cite{Gautier2022SciAd}. In addition, Sagnac interference has been realized with free electron waves in vacuum \cite{Hasselbach1993PRA}.

In 1988, Mashhoon suggested that spin--rotation coupling leads to a novel spin-rotation effect, characterized by a special phase shift \cite{Mashhoon1988PRL} (see also Ref.~\cite{Anandan:1981zg}). The existence of this coupling was confirmed in a neutron interferometry experiment \cite{Danner2020QI}, in which a rotating magnetic field was employed. The ratio of Mashhoon to Sagnac phase shift in this experiment is of the order $10^{-10}$.

It is well known that the rotational sensitivity of a matter-wave Sagnac interferometer for particles with rest mass $m$ is significantly enhanced compared to that of an optical interferometer using light of frequency $\omega$, exceeding it by a factor of \cite{Clauser1988PC}
\begin{equation}
\label{enhancement}
\vartheta = \frac{m c^2}{ \hbar \omega} = \frac{ \omega_m}{\omega} \, .
\end{equation}
Here, $\omega_m = m c^2/\hbar$ denotes the de~Broglie (or Compton) frequency of a particle with rest mass $m$. The estimate in Eq.~(\ref{enhancement}) corresponds to the ratio of the Sagnac phase shifts for matter-wave and optical interferometers that enclose the same (projected) area and rotate with the same angular velocity. The resulting enhancement factor is substantial: for atoms, it is on the order of $\vartheta \sim 10^{10}$, while for electrons it reaches approximately $10^6$.

This amplification, together with recent advancements in Aharonov--Bohm interferometry, has motivated proposals to realize the Sagnac effect in solid-state systems using arrays of mesoscopic, ring-shaped Mach--Zehnder electron interferometers, including those based on graphene \cite{Zivkovic2008PRB, Search2009PRA, Toland2010PLA, Search2011patent}. However, there is some controversy surrounding the estimate of the enhancement factor $\vartheta$ presented in Refs.~\cite{Zivkovic2008PRB, Toland2010PLA}, where the effective carrier mass $m^\ast$ is used in place of the rest mass. This issue is particularly evident in monolayer graphene, where the charge carriers exhibit a linear dispersion relation and effectively zero mass, resembling relativistic particles such as photons rather than massive ones.

In our recent work \cite{Fesh2024PRB}, we argued that the Sagnac effect in Dirac materials, despite their relativistic-like quasiparticle dispersion, is nevertheless governed by the rest mass of a free electron. In the present paper, we provide additional arguments and evidence supporting this viewpoint. In particular, we employ the Larmor theorem to demonstrate a close connection between the Sagnac and Aharonov--Bohm effects for electrons in arbitrary materials. We then apply this result to provide an alternative derivation of the Sagnac effect for electrons in graphene.

In our previous studies, we limited our analysis to the squared Dirac Hamiltonian, thereby neglecting the pseudospin degree of freedom. In this paper, we extend our approach by explicitly incorporating pseudospin as well as the electron's intrinsic spin. Within this framework, the electron pseudospinor wave function in graphene acquires an additional Berry phase factor \cite{Pichler2011Physics}, which plays a significant role in governing interference phenomena. Furthermore, we take into account the intrinsic spin of the electron and its possible splitting, which allows us to consider a graphene-based analog of the Mashhoon effect \cite{Mashhoon1988PRL}.

The paper is organized as follows. In Sec.~\ref{sec:mass}, we present a general description of the Sagnac effect in materials and argue that it is characterized by the the vacuum electronic mass in solids. In Sec.~\ref{sec:model}, we describe the elementary excitation equations of graphene, and describe our method for obtaining covariant wave equations in comoving reference frame. In Sec.~\ref{sec:relat}, we give a relativistic derivation of the Sagnac and Mashhoon effects for Dirac quasiparticles in a rotating nanotube and rotating planar ring. In Sec.~\ref{sec:nonrel}, we provide an alternative derivation of these effects based on the non-relativistic Larmor theorem. We summarize and discuss our results in Sec.~\ref{sec:conclusion}\@. Appendix~\ref{app:Dirac} provides the necessary details for deriving the Pauli equation in a rotating frame, while Appendix~\ref{app:Larmor} establishes and discusses the effective Larmor theorem.

\section{Sagnac effect for material particles: rest vs effective mass}
\label{sec:mass}

Consider a wave-like process occurring either in vacuum or within a medium and characterized by a phase $S$, which is a scalar function of space and time, that is, unambiguously defined at every point in space-time. The Sagnac effect for such a process can be derived from the following two observations:
\begin{enumerate}

\item Let a detector follow a world line $x^\mu (\tau)$, where $\tau$ is its proper time. Then the frequency of the wave measured by this detector is given by 
\begin{equation}\label{obsom}
\omega = \frac{d S \left( x ( \tau) \right)}{d \tau} = \frac{d x^\mu (\tau)}{d \tau} \nabla_\mu S = u^\mu \nabla_\mu S \, ,
\end{equation}
where $u^\mu = d x^\mu / d \tau$ denotes the four-velocity of the detector.

\item In the instantaneous {\em local inertial rest frame\/} of a material element, the phase $S$ satisfies the relation
\begin{equation}
\dot S^2/c^2 - \bigl( {\boldsymbol \nabla} S \bigr)^2 = \omega^2/c^2 - k^2 \, , 
\end{equation}
where the overdot and $\boldsymbol \nabla$ denote, respectively, the time and spatial derivatives in this frame, and $\omega$ and $k$ denote the frequency and the corresponding wave number measured in this frame. This relation can be generalized to an arbitrary coordinate frame, where it takes the covariant form
\begin{equation} \label{k-norm}
g^{\mu\nu} \nabla_\mu S \nabla_\nu S = \omega^2/c^2 - k^2 \, ,
\end{equation}
with $g^{\mu\nu}$ being the space-time metric in this frame.
\end{enumerate}

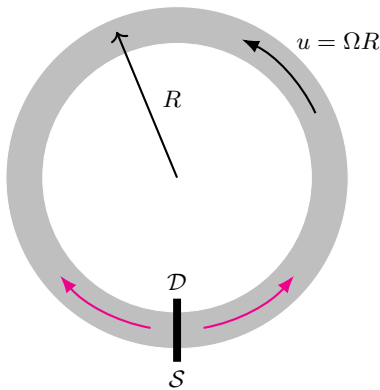
\begin{figure}[htp]
\centering
\begin{tikzpicture}[scale=.7] 
	\draw [white, fill=gray, opacity=.5] (3,3) circle [radius=3.25];
	\draw [white, fill=white] (3,3) circle [radius=2.55];
	\draw [->, thick, black] (3,3) -- ++(112.5:3); 
    \centerarc[-Latex, thick, magenta](3,3)(-80:-40:2.9);
    \centerarc[-Latex, thick, magenta](3,3)(-100:-140:2.9);
    \centerarc[-Latex, thick](3,3)(25:65:2.9);
	\draw [line width=3] (3,.7) -- (3,-.5);
	\node [] at (2.9,4.5) {$R$};
	\node [above] at (3,.7) {${\cal D}$};
	\node [below] at (3,-.5) {${\cal S}$};
	\node [right] at (5.1,5.6) {$u = \Omega R$};
\end{tikzpicture} 
\caption{{\bf Schematic of the Sagnac experiment.} Waves originating in phase at the source ${\cal S}$ propagate in opposite directions around a rotating loop, producing a measurable phase shift at the detector ${\cal D}$.}
\label{fig:Sagnac}
\end{figure}

For a simple setup of the Sagnac experiment, consider a thin ring of radius $R$, either composed of material or filled with vacuum, rotating uniformly with an angular velocity $\Omega$. A source ${\cal S}$, co-rotating with the ring, emits (or splits) waves that subsequently propagate in opposite directions along the ring’s circumference. After completing their respective paths, the waves interfere at a detector ${\cal D}$, which is also co-rotating with the system (see Fig.~\ref{fig:Sagnac}).

Let us examine the Sagnac effect from the perspective of the rotating frame. In this frame, the space-time metric on the ring world sheet can be expressed in the form:
\begin{equation} \label{metric}
d s^2 = c^2 d \tau^2 - 2 \Omega R\, \gamma d \tau d x - d x^2 \, ,
\end{equation}
where $\tau$ is the proper time of the ring material, $\gamma = \left( 1 - \Omega^2 R^2 / c^2 \right)^{-1/2}$ is the Lorentz factor associated with rotation, and $x$ is the spatial periodic coordinate along the ring with period $2 \pi R$. 

Note that the ring is made of a solid material and, therefore, should undergo Lorentz contraction when set into rotational motion. Thus, the radius $R$ refers to its value in this rotating state as measured in the laboratory frame. For a free, unsupported ring, the relation $\gamma R = R_0$ would hold, where $R_0$ is the radius of a non-rotating ring. However, if the ring is supported\,---\,as is typically the case\,---\,its deformation during rotation depends on the strain in the supporting material and may be less pronounced.

Considering two waves that start with identical phases at the position $x = 0$ and propagate in opposite directions, we seek a solution for their phases in the form
\begin{equation}
S_\pm = \omega \tau - k_\pm x \, , 
\end{equation}
where $k_+ > k_-$ are the corresponding wave vectors. Assuming, for simplicity, the dispersion relation to be isotropic in the rest frame and applying relation \eqref{k-norm} with metric \eqref{metric} to both phases $S_\pm$, we obtain the expression for the wave vectors:
\begin{equation}\label{eq:kpm}
k_\pm = \gamma \left( \Omega R \omega/c^2 \pm k \right) \, .
\end{equation}
The Sagnac fringe shift $\Theta_\text{S} = S_{-} - S_{+}$, accumulated after both waves complete one full circle around the ring, is given by
\begin{equation} \label{Sagnac}
\Theta_{\mathrm{S}} = 2 \pi R \left( k_+ + k_- \right)  = \frac{4 \pi R^2 \gamma \omega \Omega}{c^2} \, .
\end{equation}
The angular displacement of the interference fringes along the circular detection path is
\begin{equation} \label{fringe}
\phi_\text{S} = \frac{\Theta_{\mathrm{S}}}{R \left( k_- - k_+ \right)} = - \frac{2 \pi R v_\text{ph} \Omega}{c^2} \, ,
\end{equation}
where $v_\text{ph} = \omega / k$ is the phase velocity of the wave.

If the waves do not complete a full circle but instead interfere at an angle $\phi_\text{A}$ measured along the path of the wave with wave vector $k_+$, then the resulting fringe shift is given by
\begin{equation} \label{Sagnac-gen}
\Theta = k_+ R \phi_\text{A} + k_- R \left( 2 \pi - \phi_\text{A} \right) = \frac12 \Theta_\text{S} + 2 k \gamma R \left( \phi_\text{A} - \pi \right) \, .
\end{equation}
The first term in this expression depends linearly on the angular velocity $\Omega$ of rotation, while the second term is practically independent of $\Omega$ and reflects the simple geometric fact that counter-propagating waves traverse different path lengths before reaching the observation point.

A notable feature of the Sagnac fringe shift \eqref{Sagnac} or \eqref{fringe} is that it depends on the frequency of the corresponding wave measured in the rest frame of the medium. The dependence of $\Theta_{\mathrm{S}} $ on the wavenumber $k$ emerges only through the dispersion relation $\omega (k)$. For a classical wave-like process, such as a sound or electromagnetic wave, the definition of frequency is unambiguous, as it is a directly observable and measurable quantity. For quantum waves, the issue is more subtle. 
In non-relativistic quantum mechanics, the phase of a wave function is not a true scalar, as it transforms non-trivially under changes between inertial frames \cite{Landau.book3, Dieks1990AJP, Ballentine:book}. Consequently, one cannot directly substitute the non-relativistic expression $\omega = \hbar k^2 / 2m$ into \eqref{Sagnac} or \eqref{fringe} to derive the Sagnac effect for a free particle of mass $m$. 

For particles in vacuum, the resolution is straightforward: one can either apply the Galilean transformation to the phase \cite{Dieks1990AJP}, or alternatively, adopt a fully relativistic wave equation, such as the Klein--Gordon or Dirac equation, where the phase factor transforms as a true scalar. In this latter case, we have
\begin{equation}
\hbar \omega = \sqrt{m^2 c^4 + \hbar^2 k^2 c^2} = m c^2 + \frac{\hbar^2 k^2}{2m} + \ldots \, ,
\end{equation}
and the first term in this expansion dominates in the non-relativistic limit, leading to the well-known result for the Sagnac effect \cite{Anandan:1977ra, Dieks1990AJP, Hendricks1990QO, Hendricks1990QO-corr}
\begin{equation} \label{nonrel}
\Theta_{\mathrm{S}} \approx \frac{4 \pi R^2 m \Omega}{\hbar} \, , \qquad \phi_\text{S} \approx - \frac{2 \pi R m \Omega}{\hbar k} \, . 
\end{equation}

Results regarding the Sagnac effect for electrons in mesoscopic systems remain limited and controversial. In Refs.~\cite{Zivkovic2008PRB, Toland2010PLA}, it was suggested that Eqs.~\eqref{nonrel} could be applied with the effective mass $m^*$ substituted for the mass $m$.  However, this substitution is not well-founded, since the effective mass $m^*$ arises solely as a parameter in the low-energy expansion of the band structure ${\cal E} (k)$ applied in the vicinity of the Fermi level. Furthermore, there exists a broad class of Dirac materials  which have charge carriers that behave like massless relativistic particles, exhibiting a pseudo-relativistic dispersion ${\cal E} (k) \propto k$.

In this regard, it is useful to start with a relativistic field and recall the original Dirac equation describing an electron with vacuum mass $m_e$ and charge $- e$ in a crystalline lattice at rest:
\begin{equation} \label{Dirac-0}
\gamma^\mu \left(  {\rm i} \hbar \partial_\mu + \frac{e}{c} A_\mu \right) \psi - m_e c \psi = 0 \, ,
\end{equation}
where $A_\mu$ is the covector potential of the electromagnetic field of the crystal.  This equation is invariant with respect to local gauge transformations, and the observables such as the current density $\bar \psi \gamma^\mu \psi$  are gauge-invariant.  For a static crystalline lattice, there exists a unique gauge in which $A_\mu$ is static and periodic in space, vanishing at spatial infinity outside the material. 

Solutions to the Dirac equation are given by the Bloch theory and determine the one-particle electronic spectrum, hence, the dispersion relation  
\begin{equation} \label{freq}
\hbar \omega ({k}) = m_e c^2 - \varepsilon ({k}) \, .
\end{equation}
Here,  $\varepsilon ({k})$ is the binding energy of electron in a state with Bloch wave number ${k}$. Note that the phase of the wave function in the relativistic equation \eqref{Dirac-0} is a scalar, and its temporal derivative corresponds to the frequency defined in \eqref{freq}. 

For real systems at low temperatures, only electrons at the Fermi level, with the corresponding Bloch wavenumber $k_\text{F}$, will participate in the Sagnac effect. Consequently, the Sagnac effect for electrons in a crystalline lattice can be derived by inserting $k = k_\text{F}$ into the general expressions \eqref{Sagnac} and \eqref{fringe}, in view of Eq.~\eqref{freq}. In the non-relativistic limit, the result reduces to Eq.~\eqref{nonrel}, with the vacuum electron mass $m_e$ substituted for $m$. 

These conclusions may appear puzzling, as they imply that the Sagnac effect in all materials is overwhelmingly dominated by the large electron rest energy $m_e c^2$. However, this result can also be derived using the effective Larmor theorem, which establishes the equivalence between rotation and a uniform magnetic field \cite{Vignale1995PL}. The universality of the Sagnac effect for electrons in mesoscopic systems then follows from the universality of the Aharonov--Bohm effect. We will use this approach in Sec.~\ref{sec:nonrel}.

In this paper, we investigate the Sagnac effect for electrons in graphene. The unique structure of graphene introduces subtle yet significant complications: it requires treating the electron's spinor amplitudes separately on the two sublattices, $A$ and $B$, leading to the emergence of an additional degree of freedom, the pseudospin. However, the main generic features of the Sagnac effect outlined above\,---\,particularly its dependence on the vacuum electron mass\,---\,remain valid.

In the following, we consistently treat the relevant spinors as relativistic and account for the contribution of the electron mass to their phase frequency. Note that the role of particle mass in the Sagnac effect has also been discussed in \cite{Wang:2024}.

\section{Model of rotating graphene}
\label{sec:model}

\subsection{Graphene at rest}

The low-energy quasiparticles in a pure non-deformed graphene at rest are described by the the envelop wave function $\Psi_s \left( t', {\boldsymbol r}' \right)$ which satisfies the following equation:
\begin{equation}
\label{graphene-rest-frame}
\left[ \Gamma^0 \left( {\rm i} \hbar \partial_{t'} -  {\cal E}_\text{D} \right) + {\rm i} \hbar v \left(\Gamma^1 \partial_{x'} + \Gamma^2 \partial_{y'} \right)  - \Gamma^3 \Delta \right] \Psi_s \left( t', {\boldsymbol r}' \right) = 0 \, .
\end{equation}
Here, the primes denote the inertial coordinates, ${\boldsymbol r}' = \left( x', y' \right)$, and $v$ is the Fermi velocity. The $4 \times 4$ matrices $\Gamma^\mu$, $\mu = 0,1,2,3$, satisfy the anticommutation relations for the Dirac matrices and are given in the Weyl (chiral) representation
\begin{equation}
\Gamma^0 = \tau_1 \otimes \sigma_0 = {\begin{pmatrix}0&\sigma_{0} \\
\sigma_{0}&0\end{pmatrix}} \, , \qquad 
\Gamma^{i} = - {\rm i} \tau_2 \otimes \sigma_i = {\begin{pmatrix}0&-\sigma_{i} \\ \sigma_{i}&0\end{pmatrix}} \, , \qquad i = 1, 2, 3 \, ,
\end{equation}
where the Pauli matrices $\tau_i$, $\sigma_i$, $i = 1, 2, 3$, as well as the $2 \times 2$ unit matrices $\tau_0$ and $\sigma_0$, act on the valley ($\mathbf{K}_\eta$ with $\eta = \pm$)
and sublattice ($A, B$) indices, respectively, of the
four-component pseudospinor $\Psi_s^T = \left( \Psi_{+s}^T, \Psi_{-s}^T \right) =
\left( \psi_{AK_+s}, \psi_{BK_+s}, \psi_{BK_-s}, \psi_{AK_-s} \right)$.
Here, $s = \pm$ labels the intrinsic spin components.

This representation is derived from a tight-binding model  for the $2p_z\, (\pi)$ orbitals of carbon atoms on the hexagonal graphene's lattice (see, e.g., Ref.~\cite{Gusynin2007review}). We consider both massless Dirac--Weyl fermions in pristine graphene and massive Dirac fermions with a mass (gap) parameterized by $\Delta$. We remind that the matrix $\gamma^3$ for the description of graphene as well as in QED$_{2+1}$ is used for the construction of the mass term rather than for the irrelevant spatial coordinate $z'$.

The energy ${\cal E}_\text{D}$ represents the energy level of the Dirac point relative to a chosen reference point. As discussed in Sec.~\ref{sec:mass} (see also Ref.~\cite{Fesh2024PRB}), Eq.~(\ref{graphene-rest-frame}) the envelope wave function originates from the Schr\"{o}dinger equation for electrons in a crystalline lattice, which itself is derived from the fundamental Dirac equation \eqref{Dirac-0} governing electron–ion interactions in solids. This implies that the full relativistic wave functions of electrons in solids include a rapidly oscillating factor $\exp\left(- {\rm i}_{} m_e c^2 t / \hbar \right)$. Although this term is typically unobservable and often omitted, it plays a crucial role in the context of the Sagnac effect in our formalism. Therefore, when analyzing Eq.~(\ref{graphene-rest-frame}), we set [see Eq.~\eqref{freq}] 
\begin{equation} \label{ED-rel}
{\cal E}_\text{D} = m_e c^2 - \varepsilon_\text{D} \, ,
\end{equation}
where $\varepsilon_\text{D}$ is the binding energy of electron at the Dirac point. It includes both the work function and the electrostatic energy. This ensures that the frequency component is present in the solutions relevant to the Sagnac effect. The binding energy $\varepsilon_\text{D}$ is negligible compared to the electron's rest energy $m_e c^2$.

In thermodynamical considerations, one often also includes a chemical potential $\mu$ in Eq.~(\ref{graphene-rest-frame}). It characterizes the carrier imbalance, that is, the difference between the densities of electrons and holes. In graphene, the value of $\mu$ can be tuned by applying a gate voltage, allowing for control over the type of charge carriers (electrons or holes). This technique is routinely employed in experiments on Aharonov--Bohm oscillations \cite{Russo2008PRB, Huefner2009PSS, Huefner2010NJP, Smirnov2012APL, Ensslin2022NanoLett, Iwakiri2023NatCom}. The electron matter wave responsible for the Sagnac effect is of the same nature as those observed in existing electron interferometers exhibiting Aharonov--Bohm oscillations. The corresponding wave numbers are given by the Fermi wave vector $k_\text{F}$, whose magnitude is determined by $|\mu| = \hbar v k_\text{F}$ for $\Delta =0$. The combination ${\cal E}_\text{D} + \mu$ corresponds to the relativistic chemical potential \cite{Landau.book5} which differs by the rest energy from the non-relativistic one. 

By seeking a solution of Eq.~(\ref{graphene-rest-frame}) in the form $\Psi_s \left( t', {\boldsymbol r}' \right) \propto \exp \left( - {\rm i}_{} {\cal E} t'/\hbar + {\rm i}_{} {\boldsymbol k} \cdot {\boldsymbol r}' \right)$, one obtains the conventional spectrum of graphene
\begin{equation}  \label{dispersion-law}
{\cal E} (k)  = \pm \sqrt{\hbar^2 v^2 k^2 + \Delta^2} +{\cal E}_\text{D} \, .
\end{equation}
Here, ${\boldsymbol k}$ denotes the wave vector measured from the Dirac point, $\mathbf{K}_\eta$, and the signs $\pm$ correspond to the energy bands above and below this point. The energy ${\cal E} (0)$ corresponds to the energy position of the Dirac point with the gap $\Delta$ taken into account.

Given that the $\mathbf{K}_\eta$ points are decoupled in Eq.~(\ref{graphene-rest-frame}), we proceed by analyzing a single point and consider the following equation in the rest frame for the two-component pseudospinor $\psi = \Psi_{-s}$, omitting the valley and spin indices and the unit matrix $\sigma_0$:
\begin{equation}
\label{graphene-rest-frame-K}
\left[ {\rm i} \hbar \partial_{t'} -  {\cal E}_\text{D} + {\rm i} \hbar v \left( \sigma_1 \partial_{x'} + \sigma_2 \partial_{y'} \right)  - \sigma_3 \Delta \right] \psi \left( t', {\boldsymbol r}' \right) = 0 \, .
\end{equation}

Multiplying this equation from the left by the operator 
${\rm i} \hbar \partial_{t'} - {\cal E}_\text{D} - {\rm i} \hbar v \left(\sigma_1 \partial_{x'} + \sigma_2 \partial_{y'} \right) + \sigma_3 \Delta$,
we arrive at the equation
\begin{equation}
\label{rest-eq}
\left( {\rm i} \hbar \partial_{t'} - {\cal E}_\text{D} \right)^2 \psi +
v^2 \hbar^2 \nabla_{{\boldsymbol r}'}^2 \psi - \Delta^2 \psi = 0 \, ,
\end{equation}
which was the starting point of our previous work \cite{Fesh2024PRB}. In that study, the spin and pseudospin degrees of freedom of electrons in graphene were neglected. In the present work, we explicitly include these degrees of freedom and base our analysis directly on Eq.~\eqref{graphene-rest-frame-K}.

\subsection{Graphene in motion}

Thus far, we have considered graphene at rest. We now turn to the case of intrinsically undeformed graphene rotating about an axis $\ell$ in the laboratory frame with angular velocity $\Omega$. We are interested in the effective wave equation at an arbitrary point in graphene. To derive this equation, we adopt the method developed in Ref.~\cite{Fesh2024PRB}, which is based on a covariant wave equation formulated for a moving medium. 

Let us denote by $\left( e_0^\mu, e_1^\mu, e_2^\mu \right)$ the orthonormal triad rigidly connected to the material, where $c e_0^\mu = u^\mu$ is the four-velocity, and $e_1^\mu$ and $e_2^\mu$ are unit vectors oriented along the specified directions of the graphene lattice, for which our equation \eqref{graphene-rest-frame-K} is formulated. 

Note that once the triad $\left( e_0^\mu, e_1^\mu, e_2^\mu \right)$ is specified, the fourth vector $e_3^\mu$ that completes it to a full space-time tetrad is uniquely determined by the orthonormality condition
\begin{equation} \label{orthon}
g_{\mu\nu} e_a^\mu e_b^\nu = \eta_{ab} \, ,
\end{equation}
together with the requirement of a consistent tetrad orientation. Here, $g_{\mu\nu}$ is the space-time metric in arbitrary coordinates, and $\eta_{ab} = \text{diag}\, \left( 1, -1, -1, -1 \right)$ denotes the Minkowski metric. 

We assume that the properties of the internal local structure (i.e., the crystalline lattice) of the material in motion are insensitive to the small accelerations caused by its motion. This means that the space-time (orbital) part of the effective equation, expressed in terms of the comoving triad, can be obtained from Eq.~\eqref{graphene-rest-frame-K} by replacing the partial derivatives with derivatives along the corresponding comoving triad vectors. 

We have the freedom to choose a separate tetrad with respect to which the intrinsic electron spin is defined. The spin equations take their simplest form in a tetrad that is co-moving with the laboratory frame but rotates about the axis of rotation with angular velocity $\Omega$. This tetrad is described in Appendix~\ref{app:Dirac} [see Eq.~\eqref{tetrad-v}], and it will also arise naturally in Sec.~\ref{sec:nonrel}\@. Accordingly, we adopt this tetrad throughout for the description of the intrinsic spin. In this tetrad, the spin projection along the rotation axis is conserved.

The analysis of our approximation to the Dirac equation \eqref{Dirac-0} in a crystal in a rotating frame, presented in Appendix~\ref{app:Dirac}, shows that the effect of rotation on the intrinsic spin in the chosen tetrad takes a particularly simple form, identical to that for a free electron (see Eq.~\eqref{Pauli} and Ref.~\cite{Matsuo2011PRL}). This treatment of the electronic spin as that of a free electron is justified insofar as the spin remains largely decoupled from the band structure. However, to account for possible deviations from the free-electron case, we introduce an effective $g$-factor $g_\Omega$ characterizing the spin–rotation coupling. In pristine monolayer graphene, this factor is expected to be close to 1, similarly to the effective magnetic $g$-factor $g_B$, which is close to the value of 2 \cite{Menezes:2016irv, Prada:2021}. Just as the magnetic $g$-factor, it may slightly depend on the orientation of the rotation axis relative to the graphene plane. In other materials, where spin-orbit coupling or other band effects are significant, it may deviate substantially from this value. Additional aspects of this $g$-factor are discussed in Appendix~\ref{app:Larmor}, where we also derive an approximate relation $g_\Omega = g_B - 1$. 

Under all these assumptions, Eq.~\eqref{graphene-rest-frame-K} can be generalized to describe rotating graphene as
\begin{equation} \label{graphene-rotating-frame-cov}
\left[ {\rm i} \hbar u^\mu \partial_\mu - {\cal E}_\text{D} + g_\Omega \widetilde \Omega s_\ell + {\rm i} \hbar v \left(\sigma_1 e_1^\mu + \sigma_2 e_2^\mu \right) \partial_\mu  - \sigma_3 \Delta \right] \psi  = 0 \, ,
\end{equation}
where $\widetilde \Omega = d \phi / d \tau$ is the angular velocity of rotation with respect to the proper time $\tau$ of the graphene point under consideration, $g_\Omega \approx 1$ is the introduced effective $g$-factor, and $s_\ell = \hbar \sigma^\text{spin}_\ell / 2$ is the operator of the electron's intrinsic spin along the axis $\ell$ of rotation, which is expressed through the Pauli matrices $\sigma_i^\text{spin}$ acting on the intrinsic spin variable. 

We emphasize that, in Eq.~\eqref{graphene-rotating-frame-cov}, rotation couples to the intrinsic (real) spin of the electron, but not to the pseudospin associated with the sublattice degree of freedom. The pseudospinor $\psi$ is defined with respect to the frame rigidly associated with the graphene lattice. Since the components of $\psi$ represent the probability amplitudes for occupying sublattices $A$ and $B$, these amplitudes remain invariant under the motion of the graphene sheet through space. This is akin to an intrinsic spinor which, when associated with a specific tetrad, transforms as a scalar under coordinate transformations. 

The difference between intrinsic spin and pseudospin is that the latter is not associated with any spin connection, provided the graphene lattice remains undeformed, as we assume here. As a result, it is not directly coupled to physical rotation of graphene in space. Pseudospin arises from the lattice basis, and when the lattice rotates rigidly, the pseudospin basis rotates with it.

\section{Sagnac and Mashhoon effects for Dirac quasiparticles}
\label{sec:relat}

 \subsection{Rotating nanotube}
 \label{sec:tube}

As mentioned above, the case in which a magnetic field is applied parallel to the axis of a carbon nanotube\,---\,i.e., when a magnetic flux $\Phi$ threads its cross-section\,---\,was theoretically studied in Ref.~\cite{Ajiki1993JPSJ} (see also Refs.~\cite{Saito1998book, Ando2005JPSJ} for reviews). For carbon nanotubes with large diameters, the effects of graphene sheet curvature can be safely neglected. In this regime, the electronic states near the Fermi level are well described by the same low-energy model as in a flat graphene sheet, with the addition of periodic boundary conditions in the circumferential direction, defined by the chiral vector $\boldsymbol{L}$, imposed on the total wave function \cite{Saito1998book, Ando2005JPSJ}. These boundary conditions can be reformulated in terms of the envelope wave functions $\psi_{A}$ and $\psi_{B}$. The presence of magnetic flux $\Phi$ is incorporated by modifying the boundary conditions with a phase factor $\exp( 2\pi {\rm i}_{} \Phi / \Phi_0)$, leading to Aharonov--Bohm oscillations of the band gap, as experimentally  studied in \cite{Cao2004PRL}.  The single-walled carbon nanotubes used in  \cite{Cao2004PRL} had the diameter $d < \SI{2}{nm}$ and  the observed Aharonov--Bohm oscillations are dependent on their chirality.

It is instructive to begin by considering the Sagnac effect in an analogous configuration, where a nanotube rotates about its axis. This setup closely resembles the case of the Aharonov--Bohm effect in a nanotube, where the coordinate along the circumferential direction is periodic. The key distinction, however, is that the wave function need not satisfy periodic boundary conditions. Furthermore, it is assumed that the radius $R$ of the nanotube (or graphene cylinder) is sufficiently large, $R \gg d/2$, allowing one to safely neglect differences between zigzag, armchair or chiral nanotubes thereby allowing the circumferential wave vector to be treated as a continuous variable.  The situation is more akin to the analysis of Aharonov--Bohm oscillations in systems with attached leads \cite{Ihn2010book}.  In the real experiment \cite{Cao2004PRL}, it is not feasible to attach the leads along the tube's generatrix. Therefore, the following setup should be regarded as a thought experiment. Nonetheless, it effectively illustrates the fundamental characteristics of the Sagnac and Mashhoon effects.

\begin{figure}[h]
\centering
\includegraphics[width=0.35\columnwidth]{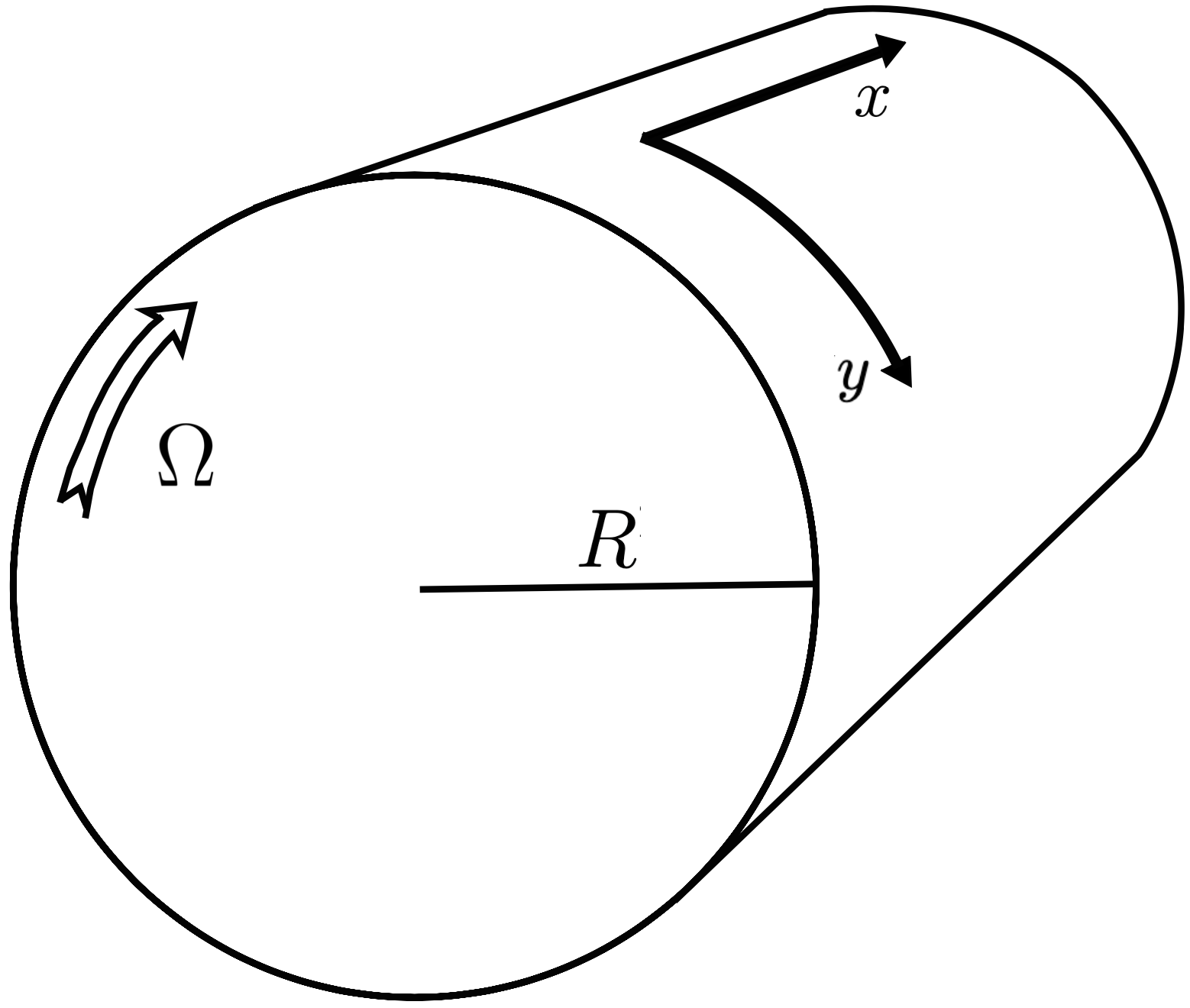}%
\caption{Comoving coordinate system $(x, y)$ in a nanotube of radius $R$ rotating around its symmetry axis with angular velocity $\Omega$ relative to the laboratory frame.}  \label{fig:tube}
\end{figure}

Consider then a carbon nanotube of radius $R$ rotating about its symmetry axis with angular velocity $\Omega$, as illustrated in Fig.~\ref{fig:tube}. We choose the local comoving coordinates $(x, y)$ on the surface of the rotating nanotube as illustrated in Fig.~\ref{fig:tube}. Thus, the $y$ coordinate runs along the circumference of the cylinder, rendering it periodic with a period of $2\pi R$. The metric interval on the graphene world hypersurface in the comoving coordinates $\left(c t, x, y \right)$ is given by
\begin{equation} \label{Galilean-interval}
d s^2 = \gamma^{-2} c^2 dt^2  - d x^2 - 2 \Omega R dt d y - d y^2 \, ,
\end{equation}
where $\gamma = \left( 1 - \Omega^2 R^2 / c^2 \right)^{-1/2}$ is the Lorentz factor. 

To write Eq.~\eqref{graphene-rotating-frame-cov} in the chosen coordinates, it remains to determine the orthonormal triad $\left( e_0^\mu, e_1^\mu, e_2^\mu \right)$. This is straightforward because the four-velocity $u^\mu$ has only a temporal component in comoving coordinates, and $e_1^\mu$ is a unit vector directed along the $x$ coordinate. The remaining vector $e_2^\mu$, tangent to the nanotube world hypersurface and pointing in the $y$ coordinate direction, is determined from the orthonormality condition \eqref{orthon}, and the triad components are 
\begin{equation}
e_0^\mu = \left( \gamma , 0, 0 \right) \, , \qquad e_1^\mu = \left( 0, 1, 0 \right) \, , \qquad e_2^\mu = \left( \gamma \Omega R/c , 0 , \gamma^{-1} \right) \, .
\end{equation}

Using these components in Eq.~\eqref{graphene-rotating-frame-cov}, we obtain
\begin{equation} \label{eq:tube}
\left[ {\rm i} \hbar \gamma \left( 1 + \frac{v \Omega R}{c^2} \sigma_2 \right)\partial_t - {\cal E}_\text{D} + \gamma g_\Omega \Omega s_1 + {\rm i} \hbar v \left(\sigma_1 \partial_x + \gamma^{-1} \sigma_2 \partial_y \right) - \sigma_3 \Delta \right] \psi  = 0 \, ,
\end{equation}
where we have taken into account that $\widetilde \Omega = \gamma \Omega$ and that the axis $x$ of rotation is used for projecting the intrinsic spin. The calculated values of the resulting effects, whether or not they include the Lorentz factor $\gamma$, are not experimentally distinguishable \cite{Post1967RMP}, due to the minuteness of $\Omega^2 R^2/c^2 \approx 10^{-21}  \left( \Omega / \text{Hz} \right)^2 \left(R / \text{cm} \right)^2$. Nevertheless, we retain the factor $\gamma$ here and in the following for the sake of formal consistency, as this factor enters the exact expressions for the metric and tetrad components.

The electronic wave is supposed to enter the nanotube at $y = 0$, where it splits into two components $\psi_\pm$ that propagate in opposite directions and later recombine after traversing the circumference of the tube. First, we consider one particular component of the intrinsic spin along the rotation axis. Then $s_1 = \text{const} = \pm \hbar/2$. 

Consider the case where the wave is not excited in the $x$ direction so that its dependence on $x$ can be neglected. Solutions for the waves can be sought for in the form
\begin{equation}\label{ans-tube}
\psi_\pm = \chi_\pm e^{- {\rm i}_{} S_\pm} \, , \quad S_\pm = \omega t / \gamma - k_\pm y \, , 
\end{equation}
where $\chi_\pm$ are constant pseudospinors, $k_+ > k_-$ are the wave vectors, and $\omega$ is the wave frequency measured in the rotating frame, according to Eq.~\eqref{obsom}. Substituting this into Eq.~\eqref{eq:tube}, we obtain an algebraic eigenvalue equation for pseudospinors $\chi_\pm$, which has a nontrivial solution if the determinant of the system vanishes. This gives the following equation for the wave vectors $k_\pm$:
\begin{equation}\label{eq:k}
\left( k_\pm - \frac{\gamma \omega \Omega R}{c^2} \right)^2 = \frac{\gamma^2}{\hbar^2 v^2} \left[ \left( \hbar \omega - {\cal E}_\text{D} + \gamma g_\Omega \Omega s_1 \right)^2 - \Delta^2 \right] \, .
\end{equation}
Its solution is 
\begin{equation}\label{kpm-tube}
k_\pm = \gamma \left[ \frac{\omega \Omega R}{c^2} \pm k_\Omega (\omega) \right]  \, ,
\end{equation}
where 
\begin{equation}\label{komega}
k_\Omega (\omega) = \frac{1}{\hbar v} \sqrt{\left( \hbar \omega - {\cal E}_\text{D} + \gamma g_\Omega \Omega s_1 \right)^2 - \Delta^2} \, .
\end{equation}

Equation \eqref{kpm-tube} reproduces the universal expression \eqref{eq:kpm} for our material and gives the usual Sagnac fringe shift characterized by Eqs.~\eqref{Sagnac}, \eqref{fringe} and \eqref{nonrel}, with electron's mass in place of $m$, if we recall that the frequency $\omega$ contains a large contribution ${\cal E}_\text{D}/\hbar \approx m_e c^2 / \hbar$.

As discussed in the Introduction, the electrons contributing to the Sagnac and Mashhoon effects are those near the Fermi surface. For such electrons, the (relativistic) frequency is given by Eq.~\eqref{freq} with $k = k_\text{F}$, and is dominated by the electron’s rest mass, yielding $\omega \approx m_e c^2 / \hbar$.

The pseudospinors $\chi_\pm$, corresponding to distinct eigenvalues of the matrix operator in Eq.~\eqref{graphene-rotating-frame-cov}, are intrinsically orthogonal. Consequently, interference between them cannot occur within the body of the nanotube. Instead, such interference arises only at the junctions where the nanotube connects to external conductors\,---\,regions where the two modes begin to propagate in parallel and are no longer pseudospin-orthogonal.

One can show that the two solutions for pseudospinor in the case $\Delta = 0$ are characterized by the helicity condition
\begin{equation}
\pm \sigma_2 \xi_{\pm} = \epsilon \chi_{\pm} \, , 
\end{equation}
where $\epsilon = \text{sign} \left( \hbar \omega - {\cal E}_\text{D} + \gamma g_\Omega \Omega s_{1}  \right)$ corresponds to electrons ($\epsilon = 1$) or holes ($\epsilon = - 1$), respectively.

Let us now take into account possible non-trivial configurations of intrinsic spin. To simplify the analysis, we set $\Delta = 0$ in what follows. The corresponding wave numbers are then given by
\begin{equation} \label{kpm}
k_\pm = \gamma \left[ \frac{\omega \Omega R}{c^2} \pm k (\omega) \pm \frac{\gamma g_\Omega \Omega s_{1}}{\hbar v} \right] \, ,
\end{equation}
where $k (\omega)$ is  the expression \eqref{komega} with $\Omega = 0$ and $\Delta = 0$.

We consider the normalized initial state to be in the superposition given by
\begin{equation}\label{psin}
\left| \psi_\text{in} \right\rangle = \left| +  \right\rangle \, + \,  \left| - \right\rangle \, ,
\end{equation}
where $\left| \pm \right\rangle$ denote the (separately unnormalized) spin eigenstates with corresponding spin projections along the rotation axis.

First, assume that the spin state remains unchanged as the two wave packets split equally upon entering the nanotube. Then, after completing the round trip along the nanotube, the corresponding wave functions, up to a common overall phase factor, take the form
\begin{equation}\label{split-tube}
\begin{split}
\left| \psi_+ \right\rangle &= \frac12 e^{{\rm i}_{} \Theta_\text{S} /2} \left( e^{{\rm i}_{} \Theta_\text{M} /2} \left| + \right\rangle \, + \, e^{- {\rm i}_{} \Theta_\text{M} / 2} \left| - \right\rangle \right) \, , \\[3pt]
\left| \psi_- \right\rangle &= \frac12 e^{- {\rm i}_{} \Theta_\text{S} /2} \left( e^{{\rm i}_{} \Theta_\text{M} / 2} \left| + \right\rangle \, + \, e^{- {\rm i}_{} \Theta_\text{M} / 2} \left| - \right\rangle \right) \, ,
\end{split}
\end{equation}
where
\begin{equation} \label{SM}
\Theta_\text{S} = \frac{4 \pi R^2 \gamma \omega \Omega}{c^2} \, , \qquad \Theta_\text{M} = \frac{2\pi R \gamma^2 g_\Omega \Omega}{v} 
\end{equation}
are, respectively, the Sagnac and Mashhoon fringe shifts. The final state is given by the superposition
\begin{equation}\label{psif}
\left| \psi_\text{fin} \right\rangle = \left| \psi_+ \right\rangle \, + \, \left| \psi_- \right\rangle = \cos \frac{\Theta_\text{S}}{2} \left( e^{{\rm i}_{} \Theta_\text{M} / 2} \left| + \right\rangle \, + \, e^{- {\rm i}_{} \Theta_\text{M} / 2} \left| - \right\rangle \right) \, .
\end{equation}

We observe that the final state exhibits both the Sagnac effect, describing the probability of detecting the electron and characterized by the fringe shift $\Theta_\text{S}$, and the Mashhoon effect of spin rotation, associated with the fringe shift $\Theta_\text{M}$. If the initial spin is in a mixed (unpolarized) state, or if the spin is not observed, we must average over the spin projections, in which case only the Sagnac effect remains observable.

Consider now a hypothetical situation in which the initial wave is split such that the components with spin $\left| \pm \right\rangle$ in the superposition \eqref{psin} propagate with wave vectors $k_\pm$, respectively \cite{Mashhoon1988PRL}. Then, after completing the round trip, the corresponding wave functions, up to a common overall phase factor, take the form
\begin{equation}
\left| \psi_+ \right\rangle = e^{{\rm i}_{} \Theta_\text{SM} /2} \left| + \right\rangle \, , \qquad 
\left| \psi_- \right\rangle = e^{- {\rm i}_{} \Theta_\text{SM} /2} \left| - \right\rangle \, ,
\end{equation}
where $\Theta_\text{SM} = \Theta_\text{S} + \Theta_\text{M}$. The final state is given by their superposition
\begin{equation} \label{psis}
\left| \psi_\text{fin} \right\rangle = \left| \psi_+ \right\rangle \, + \, \left| \psi_- \right\rangle = e^{{\rm i}_{} \Theta_\text{SM} / 2} \left| + \right\rangle \, + \, e^{- {\rm i}_{} \Theta_\text{SM} / 2} \left| - \right\rangle \, .
\end{equation}
We observe that, in this case, there is no classical Sagnac interference effect, in the sense that the probability of detecting the electron remains unity. The only observable phenomenon is the Mashhoon effect of spin rotation, characterized by the combined fringe shift $\Theta_\text{SM}$.

Note that, while the nanotube configuration considered in this section is a gedankenexperiment, the analogous setup for rotating graphene rings, considered in the next subsection, is quite realistic.  

\subsection{Rotating ring}
\label{sec:ring}

As mentioned in the Introduction, numerous interference experiments have been conducted on graphene ring structures. Consider then a thin planar ring made of graphene, rotating about its symmetry axis while occupying an average radius $R$, as shown in Fig.~\ref{fig:Sagnac}. In the polar coordinates $(ct, r, \phi)$ comoving with the ring, where $x = r \cos \phi$ and $y = r \sin \phi$, the space-time metric on the graphene hypersurface has the form
\begin{equation}\label{metric-polar}
d s^2 = \gamma^{-2} c^2 d t^2 - 2 \Omega r^2 d t d \phi - d r^2 - r^2 d \phi^2 \, ,
\end{equation} 
with the Lorentz factor $\gamma (r) = \left( 1 - \Omega^2 r^2 / c^2 \right)^{-1/2}$. 

For a moment, consider graphene at rest, i.e., the case of $\Omega = 0$. In this case, it is convenient to write Eq.~\eqref{graphene-rest-frame-K} in the polar coordinates (Ref.~\cite{Recher:2007}):
\begin{equation}
\label{graphene-rest-polar}
\left[ {\rm i} \hbar \partial_{t} - {\cal E}_\text{D} + {\rm i} \hbar v \left( \sigma_r \partial_r + \sigma_\phi r^{-1} \partial_\phi \right) - U (r) \sigma_3 \right] \psi = 0 \, .
\end{equation}
Here, we have introduced the polar Pauli matrices
\begin{equation}\label{srp}
\begin{split}
\sigma_r &= \sigma_1 \cos \phi + \sigma_2 \sin \phi = \begin{pmatrix} 0 & e^{- {\rm i}_{} \phi}  \\ e^{{\rm i}_{} \phi} & 0 \end{pmatrix} \, , \\[3pt]
\sigma_\phi &= \sigma_2 \cos \phi - \sigma_1 \sin \phi = \begin{pmatrix} 0 & - {\rm i}_{} e^{- {\rm i}_{} \phi}  \\ {\rm i}_{} e^{{\rm i}_{} \phi} & 0 \end{pmatrix} \, , 
\end{split}
\end{equation}
and replaced the constant mass parameter $\Delta$ with a mass-type potential $U(r)$, whose role is to confine the electron’s wave function to the narrow ring (see \cite{Berry:1987qi, Recher:2007, Gioia:2018}).

To obtain the wave equation in a rotating graphene in the polar coordinates of metric \eqref{metric-polar}, it is necessary to replace the orthonormal triad $\left( \partial_{ct}, \partial_r, r^{-1} \partial_\phi \right)$ present in \eqref{graphene-rest-polar} with a new orthonormal triad rigidly attached to graphene, which we denote as $\left( e_0, e_r, e_\phi \right)$, and to add a term connected with the intrinsic spin. We thus obtain [cf.\@ Eq.~\eqref{graphene-rotating-frame-cov}]
\begin{equation} \label{graphene-rotating-frame-pol}
\left[ {\rm i} \hbar c e_0^\mu \partial_\mu - {\cal E}_\text{D} + g_\Omega \widetilde \Omega s_\ell + {\rm i} \hbar v \left(\sigma_r e_r^\mu + \sigma_\phi e_\phi^\mu \right) \partial_\mu  - U \sigma_3 \right] \psi  = 0 \, .
\end{equation}

As a side remark, we note that the locally observed energy ${\cal E}_\text{D}$ and chemical potential in a static gravitational field scale with position as $1/\sqrt{g_{00}} = \gamma$ \cite[\S,27]{Landau.book5}, so they slightly depend on $r$.

In the new triad, the timelike vector corresponds to the four-velocity of the material, $e_0^\mu = u^\mu / c$; the radial vector remains unchanged; and the vector that points in the angular spatial direction is determined by the orthonormality condition using metric \eqref{metric-polar} (the corresponding tetrad was adopted, e.g., in \cite{Soares:1995cj, Strange2016PLA} for the vacuum Dirac equation in rotating frame):
\begin{equation}\label{comp}
e_0^\alpha = \left( \gamma,\, 0,\, 0 \right) \, , \qquad
e_r^\alpha = \left( 0,\, 1,\, 0 \right) \, , \qquad 
e_\phi^\alpha = \left( \frac{\gamma \Omega r}{c},\, 0,\, \gamma^{-1} r^{-1} \right) \, .
\end{equation}

Substituting these tetrad components into Eq.~\eqref{graphene-rotating-frame-pol} and taking into account that $\widetilde \Omega = \gamma \Omega$, we obtain
\begin{equation} \label{eq:ring}
\left[ {\rm i} \hbar \gamma \left( 1 + \frac{v \Omega r}{c^2} \sigma_\phi \right)\partial_t - {\cal E}_\text{D} + \gamma g_\Omega \Omega s_3 + {\rm i} \hbar v \left(\sigma_r \partial_r + \gamma^{-1} r^{-1} \sigma_\phi \partial_\phi \right) - U \sigma_3 \right] \psi  = 0 \, ,
\end{equation}
where the intrinsic spin is now projected along the $z$ axis, which is the axis of rotation.

Similarly to the ansatz \eqref{ans-tube} of the previous case, we look for stationary solutions in the rotating frame in the form
\begin{equation}\label{ans-ring}
\psi_\pm = \chi_\pm e^{- {\rm i}_{} S_\pm } \, , \qquad S_\pm = \omega_0 t - q_\pm \phi \, , 
\end{equation}
where $\omega_0$ is a constant frequency, and $q_+ > q_-$ and $\chi_\pm$ are independent of $t$. As in the case of nanotube, in the treatment of the Sagnac effect, the wave function need not satisfy periodic angular boundary conditions, and our parameters $q_\pm$ are continuous. 

Substituting \eqref{ans-ring} into \eqref{eq:ring}, we obtain equations for the pseudospinors $\chi_\pm$ for a fixed $z$-component $s_3$ of the intrinsic spin:
\begin{align}
\left( \hbar \omega - {\cal E}_\text{D} + \gamma g_\Omega \Omega s_{3} + \frac{v \hbar \omega \Omega r}{c^2} \sigma_\phi \right) \chi_\pm + {\rm i} \hbar v \left( \sigma_r \partial_r + \gamma^{-1} r^{-1} \sigma_\phi \partial_\phi \right) \chi_\pm \nonumber \\ {} - \hbar v \gamma^{-1} r^{-1} q_\pm \sigma_\phi \chi_\pm - U \sigma_3 \chi_\pm = 0 \, ,
\end{align}
where $\omega = \gamma \omega_0$ is the frequency observed in the rotating frame.

Given the structure of sigma-matrices \eqref{srp}, we look for solution in the form 
\begin{equation} \label{chi}
\chi_\pm = \begin{pmatrix} \xi_{1\pm} (r)\, e^{- {\rm i}_{} \phi/2} \\[2pt] \xi_{2\pm} (r)\, e^{{\rm i}_{} \phi/2} \end{pmatrix} \, .
\end{equation}
Then the pseudospinors
\begin{equation}
\xi_\pm = \begin{pmatrix} \xi_{1\pm} (r) \\[2pt] \xi_{2\pm} (r) \end{pmatrix}
\end{equation}
satisfy the equation
\begin{equation}\label{eqxi}
\left[ \hbar \omega - {\cal E}_\text{D} + \gamma g_\Omega 
\Omega s_{3} + \left( \frac{v \hbar \omega \Omega r}{c^2} - \frac{\hbar v}{\gamma r} q_\pm \right) \sigma_2 \right] \xi_\pm + {\rm i} \hbar v \sigma_1 \left( \partial_r + \frac{1}{2 \gamma r} \right) \xi_\pm - U \sigma_3 \xi_\pm = 0 \, .
\end{equation}

Equation \eqref{eqxi} is still exact, and we are now going to find its approximate solutions in the limit of narrow ring. For a narrow ring (with width much smaller than the wavelengths $2 \pi R / q_\pm$), we make an assumption that the radial pseudospinors $\xi_\pm (r)$ practically do not depend on the state of circular motion, i.e., on $q_\pm$, having a universal confinement profile in the narrow ring. Inspecting then Eq.~\eqref{eqxi}, we conclude that the radial pseudospinors should satisfy the equation 
\begin{equation}\label{zero-k}
{\rm i} \hbar v \sigma_1 \left( \partial_r + \frac{1}{2 \gamma r} \right) \xi_\pm - U \sigma_3 \xi_\pm = {\cal E} \xi_\pm \, , 
\end{equation}
where ${\cal E}$ is the energy eigenvalue for the confinement.

By virtue of Eq.~\eqref{zero-k}, equation \eqref{eqxi}, in the limit of an infinitesimally narrow ring of radius $R$, gives a purely algebraic relation:
\begin{equation} \label{ring-angular}
\left( \hbar \omega - \widetilde {\cal E}_\text{D}  + \gamma g_\Omega \Omega s_{3} \right) \xi_\pm + \left(\frac{v \hbar \omega \Omega R}{c^2} - \frac{\hbar v}{\gamma R} q_\pm \right) \sigma_2 \xi_\pm  = 0 \, ,
\end{equation}
where $\widetilde {\cal E}_\text{D} = {\cal E}_\text{D} - {\cal E}$, so that the energy ${\cal E}$ simply shifts the value of the energy of the Dirac point. We remember that ${\cal E}_\text{D}$ in our approach contains a large contribution from the electron's rest energy $m_e c^2$, compared to which this energy shift is quite small. 

Equations \eqref{zero-k} and \eqref{ring-angular} effectively capture the freezing of radial momentum in Eq.~\eqref{eqxi}, in a manner analogous (but not identical) to the treatment of Aharonov--Bohm oscillations a non-rotating graphene ring (see \cite{Costa:2014} and references therein).

By setting $\Omega =0$ in Eq.~\eqref{ring-angular}, it is easy to see that it describes the waves with $q_{\pm} = \pm q$ moving
counterclockwise and clockwise, respectively. In the case of graphene, these waves are protected by
the conserved helicity \cite{Gusynin2007review, Katsnelson:2020},
viz.\@ $q_\pm \sigma_{2} \xi_{\pm} \propto \left( \hbar \omega - \widetilde {\cal E}_\text{D}  \right) \xi_\pm $. In terms of the full pseudo\-spinor \eqref{chi}, this reads $q_\pm \sigma_{\phi} \chi_{\pm} \propto \left( \hbar \omega - \widetilde {\cal E}_\text{D}  \right) \chi_\pm $.  Thus, depending on the sign of $\hbar \omega - \widetilde {\cal E}_\text{D}$, we are dealing either with electrons (positive helicity) or with holes (negative helicity).

From the condition of the existence of non-trivial solutions of Eq.~\eqref{ring-angular}, we obtain
\begin{equation}\label{qpm}
q_\pm = \gamma \frac{\Omega R^2 \omega}{c^2} \pm \frac{\gamma R}{\hbar v}  \left| \hbar \omega - \widetilde {\cal E}_\text{D} + \gamma g_\Omega \Omega s_{3} \right| \, .
\end{equation}
Substituting this back into Eq.~\eqref{ring-angular}, we obtain the helicity condition
\begin{equation}\label{helical}
\pm \sigma_2 \xi_{\pm} = \epsilon \xi_{\pm} \, , \qquad 
\pm \sigma_\phi \chi_\pm = \epsilon \chi_\pm \, , 
\end{equation}
where $\epsilon = \text{sign} \left( \hbar \omega - \widetilde {\cal E}_\text{D} + \gamma g_\Omega \Omega s_{3}  \right)$ corresponds to electrons ($+ 1$) or holes ($- 1$), respectively.

Equation \eqref{qpm} reproduces the universal expression \eqref{eq:kpm} for our material and gives the usual Sagnac effect characterized by Eqs.~\eqref{Sagnac}, \eqref{fringe} and \eqref{nonrel}, with electron's mass in place of $m$, if we recall that the frequency $\omega$ contains a large contribution ${\cal E}_\text{D}/\hbar \approx m_e c^2 / \hbar$.

Similarly to the case of a nanotube, the pseudospinors $\chi_\pm$ are orthogonal on the ring, as corresponding to opposite eigenvalues of $\sigma_\phi$. As a result, interference between them cannot occur during circular propagation within the body of the ring. Such interference arises only at the junctions where the ring connects to external conductors\,---\,regions in which the two modes begin to propagate in parallel and are no longer orthogonal. As discussed in the Introduction, the corresponding Mach--Zehnder electron interferometers are routinely used to observe Aharonov--Bohm oscillations.

To give a full description of the interference effects in a ring, it is also necessary to take the phases in the pseudospinor Eq.~\eqref{chi} into account. In the context of wave propagation, the variable $\phi$ denotes the angle between the wave vector and the $y$-axis, which is rigidly aligned with the graphene crystalline lattice. 

Consider pseudospinor waves of the form \eqref{chi} propagating along the ring circumference. As we follow the waves $\chi_\pm (\phi)$ that complete a full round trip along the ring, we observe that they change sign:
\begin{equation}\label{sign-change}
\chi_\pm \left( \pm 2 \pi \right) = - \chi_\pm (0) \, .
\end{equation}
In this sense, they accumulate the overall phase $\pi$ or $- \pi$, and this phase is commonly encountered in discussions of Aharonov--Bohm effects in graphene \cite{Recher:2007, Costa:2014, Gioia:2018} and is often referred to as the Berry phase \cite{Berry:1984jv}, owing to its topological nature
\cite{Fuchs:2010}. For this reason, we will also refer to the phase in \eqref{chi} as the Berry phase. Due to the $\pi$-phase shift, conductance minima in Aharonov--Bohm oscillations appear at integer values of $\Phi/\Phi_0$ in Dirac materials, in contrast to conventional systems with Schr{\"o}dinger-type carriers, where they occur at half-integer values \cite{Zhang:2010:PRL, Gioia:2018}.  These manifestations of the Berry phase have indeed been observed in quasi-ballistic three-dimensional topological insulator nanowire devices that are gate-tunable through the Dirac point \cite{Cho:2015-nat}.

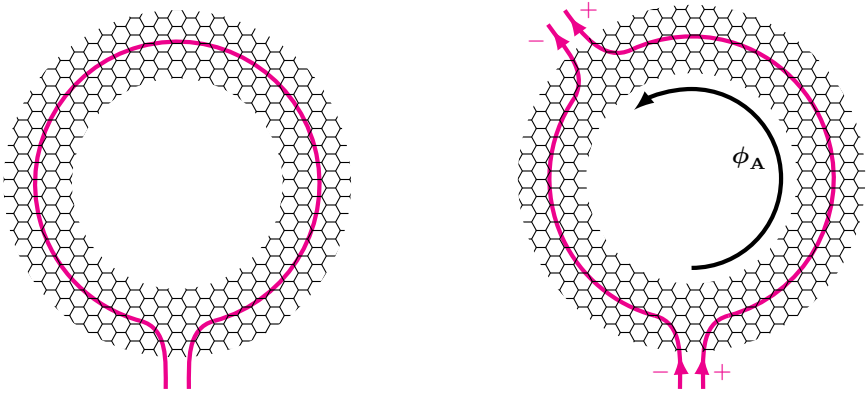
\begin{figure*}[htp]
\centering
\begin{tikzpicture}[scale=.7] 
\draw[ultra thick, magenta]  (-0.699,-2.608) to [in=90,out=-15] (-0.22,-3.9);
\draw[ultra thick, magenta] (0.22,-3.9) to [out=90,in=195] (0.699,-2.608);
\centerarc[ultra thick, magenta](0,0)(-75:255:2.7);
\draw [draw=none, pattern=hexagons] (0,0) circle [radius=3.3];
\draw [draw=none, fill=white] (0,0) circle [radius=2];
\end{tikzpicture} \hspace{2cm}
\begin{tikzpicture}[scale=.7] 
\draw[->-=.41, ultra thick, magenta]  (-0.22, -4) to [out=90,in=-15] (-0.699,-2.608);
\draw[->-=.41, ultra thick, magenta] (0.22, -4) to [out=90,in=195] (0.699,-2.608);
\draw[->-=.92, ultra thick, magenta] (-1.141, 2.447) to [out=205,in=-55] (-2.407, 3.195);
\draw[->-=.92, ultra thick, magenta] (-2.212, 1.549) to [out=65,in=-55] (-2.728, 2.925);
\centerarc[ultra thick, magenta](0,0)(-75:115:2.7);
\centerarc[ultra thick, magenta](0,0)(145:255:2.7);
\draw [draw=none, pattern=hexagons] (0,0) circle [radius=3.3];
\draw [draw=none, fill=white] (0,0) circle [radius=2];
\centerarc[-latex, ultra thick](0,0)(-90:130:1.7);
\node [] at (1.1,.4) {$\boldsymbol{\phi}_\text{\bf A}$};
\node [above right, magenta] at (0.24, -4) {$+$};
\node [above left, magenta] at (-0.24, -4) {$-$};
\node [right, magenta] at (-2.3, 3.195) {$+$};
\node [below left, magenta] at (-2.6, 2.925) {$-$};
\end{tikzpicture}
\caption{{\bf Illustration of the Berry phase in a graphene ring.} {\em Left:\/} Waves propagating counterclockwise ($+$) and clockwise ($-$) around the full ring acquire Berry phases $\phi_\pm = \pm \pi$ between entrance and exit. {\em Right:\/} For waves with entrance and exit separated by an angular distance $\phi_\text{A}$, both propagation directions acquire the same Berry phase: $\phi_\pm = \phi_\text{A} - \pi$.}
\label{fig:Berry}
\end{figure*}

Since both pseudospinors $\chi_\pm$ undergo a sign change \eqref{sign-change} relative to their starting values, no relative phase difference arises between the two waves from the round trip alone. Instead, the relative phase factors between the waves emerge from the contribution of the points of entry and exit from the ring, as illustrated in Fig.~\ref{fig:Berry}. The wave $\chi_+$, propagating anticlockwise, acquires an additional angular parameter change of $\Delta \phi_+ = -\pi$ upon entering and exiting the ring, whereas the wave $\chi_-$, propagating clockwise, experiences an angular parameter change of $\Delta \phi_- = \pi$.

Therefore, the wave $\chi_+$, propagating anticlockwise, acquires the total angular parameter change $\phi_+ = 2 \pi -\pi = \pi$ between entering and exiting the ring, whereas the wave $\chi_-$, propagating clockwise, acquires the total change of $\phi_- = - 2 \pi + \pi = -\pi$. This angular parameter change is constant and does not depend on the state of rotation.

To summarize, we consider the two waves $\chi_\pm$, which appear at the entrance to the ring in the common pseudospin state: $\chi_\pm = \left( \xi_1, \xi_2 \right)^\intercal$ (this is actually just one wave before splitting). According to the reasoning above, at the exit of the ring, their pseudospin states take the form
\begin{equation}
\chi_\pm = \begin{pmatrix} \xi_1\, e^{- {\rm i}_{} \phi_\pm /2} \\[2pt] \xi_2\, e^{{\rm i}_{} \phi_\pm/2} \end{pmatrix} \, ,
\end{equation}
where $\phi_\pm = \pm \pi$. Thus, at the exit, we observe the relation $\chi_- = e^{ {\rm i} \pi} \chi_+$, indicating a phase difference of $\pi$ between the two waves. This constant phase difference should be taken into account when calculating the Sagnac phase shift.

The case of wave propagation along a planar ring involves an additional  Berry phase because the wave vector $\boldsymbol{k}$ forms a varying angle with the underlying crystalline structure of the graphene lattice. In contrast, for the case of nanotube discussed in Sec.~\ref{sec:tube}, this angle remains constant throughout propagation, and therefore no Berry phase is present in Eq.~\eqref{ans-tube}.

The initial and final states of the intrinsic spin in the case of propagation along a ring are described by expressions quite analogous to Eqs.~\eqref{psin} and \eqref{psif}, \eqref{psis}, but referred to the spin projection $s_{3}$ and incorporating a contribution due to the Berry phase.

Specifically, in the case of no spin splitting, the initial state \eqref{psin}, after the propagation around the circle, becomes (up to a common irrelevant phase factor)
\begin{align} \label{no-split}
\left| \psi_+ \right\rangle &= \frac12 e^{{\rm i}_{} \left( \Theta_\text{S} - \pi \right) /2} \left( e^{{\rm i}_{} \Theta_\text{M} /2} \left| + \right\rangle \, + \, e^{- {\rm i}_{} \Theta_\text{M} / 2} \left| - \right\rangle \right) \, , \nonumber \\[-6pt] & \\[-6pt] 
\left| \psi_- \right\rangle &= \frac12 e^{- {\rm i}_{} \left( \Theta_\text{S} - \pi \right) /2} \left( e^{{\rm i}_{} \Theta_\text{M} / 2} \left| + \right\rangle \, + \, e^{- {\rm i}_{} \Theta_\text{M} / 2} \left| - \right\rangle \right) \, , \nonumber
\end{align}
where the additional $\pi$ in the phases is due to the Berry phase difference, and $\Theta_\text{S}$ and $\Theta_\text{M}$ are given by Eq.~\eqref{SM}. The final state is given by the superposition
\begin{align}
\left| \psi_\text{fin} \right\rangle = \left| \psi_+ \right\rangle \, + \, \left| \psi_- \right\rangle &= \cos \frac{\Theta_\text{S} - \pi}{2} \left( e^{{\rm i}_{} \Theta_\text{M} / 2} \left| + \right\rangle \, + \, e^{- {\rm i}_{} \Theta_\text{M} / 2} \left| - \right\rangle \right) \nonumber \\ &= \sin \frac{\Theta_\text{S}}{2} \left( e^{{\rm i}_{} \Theta_\text{M} / 2} \left| + \right\rangle \, + \, e^{- {\rm i}_{} \Theta_\text{M} / 2} \left| - \right\rangle \right) \, .
\end{align}
This represents the Sagnac effect, with the fringe shifted by $\pi$ relative to the previous case of nanotube, along with the usual Mashhoon effect of spin rotation.

In the case of spin splitting, the components with spin $\left| \pm \right\rangle$ in the superposition \eqref{psin} propagate with wave vectors $q_\pm$, respectively. Then, after completing the round trip, the corresponding wave functions, up to a common overall phase factor, take the form
\begin{equation}
\left| \psi_+ \right\rangle = e^{{\rm i}_{} \left( \Theta_\text{SM} - \pi \right) /2} \left| + \right\rangle \, , \qquad
\left| \psi_- \right\rangle = e^{- {\rm i}_{} \left( \Theta_\text{SM} - \pi \right) /2} \left| - \right\rangle \, ,
\end{equation}
where $\Theta_\text{SM} = \Theta_\text{S} + \Theta_\text{M}$. The final state is given by their superposition
\begin{equation} \label{fin-split}
\left| \psi_\text{fin} \right\rangle = \left| \psi_+ \right\rangle \, + \, \left| \psi_- \right\rangle = e^{{\rm i}_{} \left( \Theta_\text{SM} - \pi \right) / 2} \left| + \right\rangle \, + \, e^{- {\rm i}_{} \left( \Theta_\text{SM} - \pi \right) / 2} \left| - \right\rangle \, .
\end{equation}
Again, we observe that there is no classical Sagnac interference effect in this case, but only the Mashhoon effect of spin rotation, characterized by the combined fringe shift $\Theta_\text{SM} - \pi$.

\subsection{Setup with two contacts}
\label{sec:realistic}

To facilitate comparison with the conventional treatment of the Sagnac and Mashhoon effects, we have thus far considered a hypothetical setup in which waves propagating in opposite directions complete exactly one full loop around the material's circumference. This configuration corresponds to that of a classical Sagnac interferometer. In a different setup, the contacts are placed at different angular positions around the graphene ring, with angular distance  $\phi_\text{A}$ between them, as shown in the right image of Fig.~\ref{fig:Berry}. This corresponds to the aforementioned Mach--Zehnder electron interferometers, where the input and output of the ring are usually located directly opposite each other, with $\phi_\text{A} = \pi$. Such interferometers are widely used to study Aharonov--Bohm oscillations and have been proposed as a platform for realizing the Sagnac effect in graphene-based devices \cite{Zivkovic2008PRB, Search2009PRA, Toland2010PLA, Search2011patent}.

In the general case, for a wave propagating counterclockwise, the total Berry phase is given by $\phi_+ = \phi_\text{A} - \pi$. For a wave propagating clockwise, it becomes $\phi_- = \pi - \left( 2\pi - \phi_\text{A} \right) = \phi_+$. Thus, in this configuration, both waves accumulate the same Berry phase. The Sagnac and Mashhoon effects, under these conditions, are described as follows. 

In the case of no splitting of intrinsic spin, the initial state \eqref{psin}, propagated along different paths, at the exit point of the ring with angular position $\phi_\text{A}$ becomes, up to a common phase factor $\exp \left[ {\rm i}_{} \Theta_\text{S} \left( \phi_\text{A} - \pi \right) / 4 \pi \right]$,
\begin{align}
\left| \psi_+ \right\rangle &= \frac12 e^{{\rm i}_{} \Theta_\text{S} / 4} \left[ \exp \left( {\rm i}_{} \left| q + \frac{\Theta_\text{M}}{4 \pi} \right|\, \phi_\text{A} \right) \left| + \right\rangle \, + \, \exp \left({\rm i}_{} \left| q - \frac{\Theta_\text{M}}{4 \pi} \right|\, \phi_\text{A} \right) \left| - \right\rangle \right] \, , \nonumber \\[-12pt] & \\
\left| \psi_- \right\rangle &= \frac12 e^{- {\rm i}_{} \Theta_\text{S} / 4} \left[ \exp \left( - {\rm i}_{} \left| q + \frac{\Theta_\text{M}}{4 \pi} \right|\, \left( \phi_\text{A} - 2 \pi \right) \right) \left| + \right\rangle \, + \, \exp \left( - {\rm i}_{} \left| q - \frac{\Theta_\text{M}}{4 \pi} \right|\, \left( \phi_\text{A} - 2 \pi \right) \right) \left| - \right\rangle \right] \, , \qquad \nonumber
\end{align}
where the common Berry phase was omitted, and the Sagnac and Mashhoon fringe shifts $\Theta_\text{S}$ and $\Theta_\text{M}$, respectively, are given by Eq.~\eqref{SM}. In deriving this equation, we have used Eq.~\eqref{qpm} and made the notation
\begin{equation}
q = \frac{\gamma R}{\hbar v} \left( \hbar \omega - \widetilde {\cal E}_\text{D} \right) \end{equation}
for the central angular momentum of the waves present in Eq.~\eqref{qpm}. The final state is given by the superposition $\left| \psi_\text{fin} \right\rangle = \left| \psi_+ \right\rangle + \left| \psi_- \right\rangle$: 
\begin{align}\label{no-split-gen}
\left| \psi_\text{fin} \right\rangle &=  e^{{\rm i}_{} \left| \pi q + {\Theta_\text{M}}/{4} \right|} \cos \left( \frac{\Theta_\text{S}}{4} + \left| q + \frac{\Theta_\text{M}}{4 \pi} \right| \left( \phi_\text{A} - \pi \right) \right) \left| + \right\rangle \nonumber \\ &\quad {} + \, 
e^{{\rm i}_{} \left| \pi q - {\Theta_\text{M}}/{4} \right|} \cos \left( \frac{\Theta_\text{S}}{4} + \left| q - \frac{\Theta_\text{M}}{4 \pi} \right| \left( \phi_\text{A} - \pi \right) \right) \left| - \right\rangle \, .
\end{align}
The spin-up and spin-down components differ only by the sign in front of $\Theta_\text{M}$. 

For the pure Sagnac effect, one should average over the spin. We then use the identity $\cos^2 \left( a + b \right) + \cos^2 \left( a - b \right) = 1 + \cos 2 a \cos 2 b $ to calculate the electric current $J \propto \left| \psi_\text{fin} \right|^2$. 

In the case $|q| \geq \Theta_\text{M} / 4 \pi$, we obtain [cf.\@ Eq.~\eqref{Sagnac-gen} noting that $q = k \gamma R$]
\begin{equation}
J \propto 1 + \cos \left[ \frac{\Theta_\text{S}}{2} + 2 |q| \left( \phi_\text{A} - \pi \right) \right] \cos \left[ \frac{\Theta_\text{M}}{2 \pi} \left( \phi_\text{A} - \pi \right) \right] \, .
\end{equation}

In the case $|q| \leq \Theta_\text{M} / 4 \pi$, we have
\begin{equation}
J \propto 1 + \cos \left[ \frac{\Theta_\text{S}}{2} + \frac{\Theta_\text{M}}{2 \pi} \left( \phi_\text{A} - \pi \right) \right] \cos \left[ 2 q \left( \phi_\text{A} - \pi \right) \right] \, .
\end{equation}

All these expressions simplify in the case of the Mach--Zehnder interferometer when $\phi_\text{A} = \pi$, and become similar to the result \eqref{psif}. In the case of spin splitting (Mashhoon experiment), the components with spin $\left| \pm \right\rangle$ in the superposition \eqref{psin} propagate with angular momenta $q_\pm$, respectively. Then, after reaching the output, the corresponding wave functions, up to a common overall phase factor, take the form
\begin{equation}
\left| \psi_+ \right\rangle = e^{{\rm i}_{} \Theta} \left| + \right\rangle \, , \qquad \left| \psi_- \right\rangle = e^{- {\rm i}_{} \Theta } \left| - \right\rangle \, ,
\end{equation}
where, depending on the value of $q$, we have 
\begin{equation}
\Theta = 
\begin{cases}
\dfrac{1}{4} \left( \Theta_\text{S} + \dfrac{q}{|q|} \Theta_\text{M} \right) + q \left( \phi_\text{A} - \pi \right)\, , &|q| \geq \dfrac{\Theta_\text{M}}{4\pi}\, , \\[10pt] 
\dfrac{1}{4} \left( \Theta_\text{S} - \Theta_\text{M} \right) + \dfrac{\Theta_\text{M}}{4 \pi} \phi_\text{A} + q \pi \, , &|q| \leq \dfrac{\Theta_\text{M}}{4\pi}\, , 
\end{cases}
\end{equation}
The final state is given by their superposition
\begin{equation} 
\left| \psi_\text{fin} \right\rangle = \left| \psi_+ \right\rangle \, + \, \left| \psi_- \right\rangle = e^{{\rm i}_{} \Theta} \left| + \right\rangle \, + \,e^{- {\rm i}_{} \Theta} \left| - \right\rangle \, .
\end{equation}
As before, we observe that there is no classical Sagnac interference effect in this case, but only the Mashhoon effect of spin rotation.

\section{Derivation based on the Larmor theorem}
\label{sec:nonrel}

In Appendix~\ref{app:Larmor}, we reproduced the Larmor theorem \cite{Vignale1995PL}, demonstrating the (approximate) mapping between the electron wave equation in a system rotating about some axis and the wave equation in the system at rest but subject to a uniform external magnetic field along the same axis. The relation between the angular velocity of rotation and the corresponding magnetic field is given by 
\begin{equation}\label{B-eff}
B = - \frac{2 m_e c}{e} \Omega \, ,
\end{equation} 
where $- e$ is the electron charge, and $m_e$ is the {\em vacuum\/} electron mass. The mapping also includes an additional spin--rotation term, which is the last term in the Pauli equation \eqref{eff-fin}.

For matter waves in vacuum, the Larmor theorem was previously employed in~\cite{Sakurai:1980te, Hendricks1990QO, Rizzi:2003uc} in the context of the Sagnac effect. It is also worth noting that the Larmor theorem has been invoked to argue that the vacuum electron mass appears in the expression for the London moment\,---\,the magnetic moment acquired by a rotating superconductor \cite{Alben1969PLA} (see also Ref.~\cite{Hirsch2014PS}).

The Larmor theorem yields a universal prescription for calculating the electronic Sagnac and Mashhoon effects in arbitrary systems: one first formulates an effective electronic Hamiltonian for the system at rest in a uniform magnetic field $B$, then replaces the magnetic field with the effective field given by Eq.~\eqref{B-eff}, and finally adds an additional spin term, $\Omega s_\ell$, to the Hamiltonian. Eventually, the spin--rotation coupling should be renormalized by the effective $g$-factor $g_\Omega$.

Using this mapping, we will repeat the calculation of the Sagnac and Mashhoon effects for a rotating ring and a nanotube in graphene to demonstrate consistency with the results obtained in Sec.~\ref{sec:relat}. This offers an alternative perspective on the emergence of the vacuum electron mass in this effect, attributing it to its role in the Larmor equivalence expressed in Eq.~\eqref{B-eff}.

Our starting point is the one-electron equation for graphene at rest in a uniform magnetic field, a generalization of Eq.~\eqref{graphene-rest-frame-K}, which takes the form
\begin{equation} \label{start}
\left( {\rm i} \hbar \partial_{t} - {\cal E}_\text{D} + v \left[ \sigma_1 \left( {\rm i} \hbar \partial_{x} + \frac{e}{c} A_1 \right) + \sigma_2 \left( {\rm i} \hbar \partial_{y} + \frac{e}{c} A_2 \right) \right] - \frac{e}{m_e c} B s_\ell - \Delta \sigma_3 \right) \psi = 0 \, , 
\end{equation}
where the electron's spin is projected to the axis $\ell$ aligned with the magnetic field.

\subsection{Rotating nanotube}

Our nanotube rotates about the $x$ axis, as illustrated in Fig.~\ref{fig:tube}. The equivalent uniform magnetic field should be aligned along the same axis. Consequently, the only nonvanishing component of its covector potential in the basis of unit vectors $\bigl( \hat x, \hat \phi, \hat r \bigr)$ in polar coordinates can be chosen as $A_\phi = - B r / 2$. At the surface of the nanotube, the only non-vanishing component of the vector potential is then
\begin{equation}
A_2 = - \frac12 B R \, .
\end{equation}
Equation \eqref{start} then reads
\begin{equation}
\left( {\rm i} \hbar \partial_{t} - {\cal E}_\text{D} + v \left[ \sigma_1 {\rm i} \hbar \partial_{x} + \sigma_2 \left( {\rm i} \hbar \partial_{y} - \frac{e}{2 c} B R \right) \right] - \frac{e}{m_e c} B s_1 - \Delta \sigma_3 \right) \psi = 0 \, .
\end{equation}

Making the substitution \eqref{B-eff}, taking into account the last spin term in \eqref{eff-fin}, and correcting the spin--rotation coupling by the effective $g$-factor, $g_\Omega$, we obtain the effective description of an electron in a rotating graphene nanotube:
\begin{equation} \label{eq:tube-f}
\left( {\rm i} \hbar \partial_{t} - {\cal E}_\text{D} + g_\Omega \Omega s_1 
+ v \bigl[ \sigma_1 {\rm i} \hbar \partial_{x} + \sigma_2 \left( {\rm i} \hbar \partial_{y} + m_e \Omega R \right) \bigr] - \Delta \sigma_3 \right) \psi = 0 \, .
\end{equation}

We compare this equation with Eq.~\eqref{eq:tube} from the relativistic approach. Apart from the appearance of the Lorentz factors $\gamma$ in \eqref{eq:tube}, the only difference between the two equations is the presence of the term with $m_e \Omega R$ in \eqref{eq:tube-f}, which replaces the operator ${\rm i} \hbar \gamma \Omega R \partial_t / c^2$ in \eqref{eq:tube}. When acting on the wave function \eqref{ans-tube}, this operator yields the factor $\hbar \omega \Omega R/ c^2$. 

Therefore, all solutions of Eq.~\eqref{eq:tube-f} can be directly obtained from those in Sec.~\ref{sec:tube} by replacing the combination $\omega \Omega / c^2$ with $m_e \Omega / \hbar$ and setting the Lorentz factor everywhere to unity. In particular, the wave vectors \eqref{kpm} will be given by 
\begin{equation}
k_\pm = \frac{m_e \Omega R}{\hbar} \pm k (\omega) \pm \frac{g_\Omega \Omega s_{1}}{\hbar v} \, ,
\end{equation}
and the Sagnac and Mashhoon fringe shifts are given by
\begin{equation} \label{SM-L}
\Theta_\text{S} = \frac{4 \pi R^2 m_e \Omega}{\hbar} \, , \qquad \Theta_\text{M} = \frac{2\pi R g_\Omega \Omega}{v} \, .
\end{equation}
The interpretation of this result is, of course, identical to that presented at the end of Sec.~\ref{sec:tube}.

\subsection{Rotating ring}

Our ring now lies in a plane perpendicular to the $z$ axis, which is the direction of a homogeneous magnetic field $B$ with covector potential given by Eq.~\eqref{Acomp}. In this case, the effective equation for the graphene pseudospinor \eqref{start}, with the electron's intrinsic spin taken into account and with $\Delta$ replaced by the confining potential $U$, reads
\begin{equation} \label{start1}
\left( {\rm i} \hbar \partial_{t} - {\cal E}_\text{D} + v \left[ \sigma_1 \left( {\rm i} \hbar \partial_{x} + \frac{e}{2 c} B y \right) + \sigma_2 \left( {\rm i} \hbar \partial_{y} - \frac{e}{2 c} B x \right) \right]  - \frac{e}{m_e c} B s_3 - U \sigma_3 \right) \psi = 0 \, .
\end{equation}
Proceeding to the polar coordinates by the substitution $x = r \cos \phi$, $y = r \sin \phi$, we write Eq.~\eqref{start1} in the form
\begin{equation} 
\label{ring-Larmor}
\left( {\rm i} \hbar \partial_{t} - {\cal E}_\text{D} + v \left[ {\rm i} \hbar \left( \sigma_r \partial_r + \sigma_\phi r^{-1} \partial_\phi \right) - \frac{e}{2 c} B r \sigma_\phi  \right] - \frac{e}{m_e c} B s_3 - U \sigma_3 \right) \psi = 0 \, ,
\end{equation}
where the polar Pauli matrices $\sigma_r$ and $\sigma_\phi$ are defined in Eq.~\eqref{srp}. As a side note, we point out that that the Landau levels obtained from Eq.~(\ref{ring-Larmor}) exhibit a relativistic-like nature, characterized by the energy scale $\sqrt{\hbar v^2 |eB|}/c$, rather than the conventional scale for free nonrelativistic electrons, $\hbar |eB|/m_e c$.

Making the substitution \eqref{B-eff}, taking into account the last spin term in \eqref{eff-fin}, and finally correcting for the effective $g$-factor, we obtain the effective description of an electron in a rotating ring:
\begin{equation} \label{eq:ring-f}
\left[ {\rm i} \hbar \partial_{t} - {\cal E}_\text{D} + g_\Omega \Omega s_3 + {\rm i} \hbar v \left( \sigma_r \partial_r + \sigma_\phi r^{-1} \partial_\phi \right) + v m_e \Omega r \sigma_\phi - U \sigma_3 \right] \psi = 0 \, .
\end{equation}

We compare this equation with Eq.~\eqref{eq:ring} from the relativistic approach. Again, apart from the appearance of the Lorentz factors $\gamma$ in Eq.~\eqref{eq:ring}, the only difference between the two equations is the presence of the term with $m_e \Omega r$ in Eq.~\eqref{eq:ring-f}, which replaces the operator ${\rm i} \hbar \gamma \Omega r \partial_t / c^2$ in \eqref{eq:ring}. When acting on the wave function \eqref{ans-ring}, this operator yields the factor $\hbar \omega \Omega R/ c^2$. 

Therefore, all solutions of Eq.~\eqref{eq:ring-f} can be directly obtained from those in Sec.~\ref{sec:ring} by replacing the combination $\omega \Omega / c^2$ with $m_e \Omega / \hbar$ and setting the Lorentz factor everywhere to unity. In particular, the wave vectors \eqref{qpm} will be given by 
\begin{equation}
q_\pm = \frac{\Omega R^2 m_e}{\hbar} \pm \frac{R \left( \hbar \omega - {\cal E}_\text{D} + g_\Omega \Omega s_{3} \right)}{\hbar v} \, ,
\end{equation}
and the Sagnac and Mashhoon fringe shifts by Eq.~\eqref{SM-L}. The interpretation of this result is identical to that presented at the end of Sec.~\ref{sec:ring}.

We conclude this section with an important remark. In the relativistic framework developed in Secs.~\ref{sec:model} and \ref{sec:relat}, it was essential to define the Dirac point energy ${\cal E}_\text{D}$ relative to the electron's vacuum energy, specifically, by including the electron rest mass, as expressed in Eq.~\eqref{ED-rel}. This requirement stems from the relativistic treatment and ensures that the spinor phase transforms as a true space-time scalar.

In contrast, the effective non-relativistic equations \eqref{eq:tube-f} and \eqref{eq:ring-f}, derived using the Larmor theorem, inherently incorporate these relativistic elements through the explicit appearance of the electron mass $m_e$ in the equivalence relation \eqref{B-eff}. In this non-relativistic context, ${\cal E}_\text{D}$ can be referenced to an arbitrary energy zero, as is customary in condensed-matter physics. Notably, in the dispersion relation \eqref{komega}, which is common to both approaches, the frequency and the Dirac-point energy enter only through their difference, $\hbar \omega - {\cal E}_\text{D}$, allowing both quantities to be shifted simultaneously by an arbitrary constant without affecting this expression.

As a consequence, the Sagnac fringe shift $\Theta_\text{S}$ given by Eq.~\eqref{SM-L}, in which the electron's vacuum mass $m_e$ appears explicitly, remains invariant under such a shift. By contrast, the relativistic expression in Eq.~\eqref{SM} involves the frequency $\omega$ and therefore does not possess this invariance.

\section{Conclusion}
\label{sec:conclusion}

In this paper, we extend our previous analysis of the effects of spatial rotation in graphene \cite{Fesh2024PRB} to include both the pseudospin and the intrinsic spin of the propagating electron. Within this extended framework, in addition to the standard Sagnac effect, we also account for a Berry phase that is constant and independent of rotation. Furthermore, the intrinsic spin of the electron and its possible splitting give rise to a graphene-based analog of the Mashhoon effect.

Building upon our earlier work, we presented more detailed arguments supporting the conclusion that the Sagnac fringe shift in graphene is determined by the vacuum mass of the electron. Our first argument is based on the relativistic phase of the electron's wave function, which transforms as a scalar and necessarily includes a dominant contribution from the electron's rest energy. This rest energy then contributes to the Sagnac fringe shift via an effective Lorentz transformation. As a second argument, we invoke the effective Larmor theorem, which establishes the equivalence between rotational motion and a uniform magnetic field, with the proportionality constant between the angular velocity of rotation and the corresponding magnetic field involving the vacuum electron mass. 

The Mashhoon fringe shift, which characterizes the dynamics of intrinsic spin, retains its standard form in graphene, with its dependence on the Fermi velocity appearing in the usual way. The expressions for the Sagnac and Mashhoon fringe shifts are presented in Eq.~\eqref{SM-L}.

In analyzing the Mashhoon effect, we have neglected the potential kinematic contribution arising from the Thomas precession of the electron's spin due to its circular motion in a ring or nanotube. Similar to what occurs in neutron interferometry \cite{Mashhoon1988PRL}, this effect is likely to contribute a constant (i.e., $\Omega$-independent) term to the Mashhoon fringe shift $\Theta_\text{M}$, on the order of $(v/c)^2 \sim 10^{-5}$. This issue requires special analysis \cite{Malykin:2006}.

We examined the Sagnac and Mashhoon effects in two systems. The first is a long nanotube rotating about its axis, considered as a thought experiment. The second is a rotating ring, assumed for simplicity to be infinitesimally narrow. This latter configuration offers a practical realization, as Aharonov–Bohm oscillations have been extensively studied in graphene rings. The descriptions of the Sagnac and Mashhoon effects are quite similar in both cases. However, the ring geometry exhibits an additional nontrivial contribution from the Berry phase. This stems from the fact that, in a ring, the wave vector of the propagating electron forms a varying angle with the underlying crystalline structure of the graphene lattice, whereas in a nanotube, this angle remains constant throughout propagation.

In the present work, we restrict ourselves to the simplest analytical treatment of the continuum model of a graphene ring with frozen radial motion. This allowed us to reveal the main features of the Sagnac and Mashhoon effects in graphene. More subtle features associated with the role of edges, their geometry, disorder, mixture of types of edge termination, leading to possible intervalley coupling and affecting pseudospin, require further investigation, perhaps, on the basis of numerous existing studies; see, e.g., Refs.~\cite{Recher:2007,Schelter2012SSC,Costa:2014,Gioia:2018}. This will definitely become necessary for practical realization of the Sagnac effect in graphene interferometers.

Let us now discuss the conditions required for observing the Sagnac and Mashhoon effects in solid-state interferometers. It is convenient to express the Sagnac fringe shift from Eq.~(\ref{SM-L}) in terms of the Compton wavelength, $\lambda_\mathrm{C} = 2 \pi \hbar / m_e c \approx \SI{0.0243}{\text{\AA}}$, yielding the estimate:
\begin{equation}
\label{Sagnac-estimate}
\Theta_\text{S} = \frac{8 \pi^2 R^2  \Omega}{c \lambda_\mathrm{C} } \approx 10^{-7} \left( \frac{R}{\SI{}{\micro\metre}} \right)^2 
\frac{\Omega}{\SI{}{Hz}} \, .
\end{equation}

The Larmor theorem allows one to map the results obtained for Aharonov--Bohm oscillations by simply replacing the corresponding oscillatory expression $\exp \left( 2 \pi {\rm i}_{} \Phi/\Phi_0 \right)$ with the expression $\exp \left( 2 \pi {\rm i}_{} \Phi_\Omega/\Phi_0 \right)$, where the effective rotational flux is given by
\begin{equation}
\Phi_\Omega = \frac{2 \pi R^2 c\, m_e  \Omega}{e} \, .
\end{equation}
The resulting phase $ 2 \pi\, \Phi_\Omega / \Phi_0$ corresponds to the rotational (Sagnac) phase accumulated by an electron traveling around the ring \cite{Vignale1995PL}.

However, the smallness of the estimate in Eq.~(\ref{Sagnac-estimate}), being a consequence of the extremely weak effective Larmor magnetic field, 
\begin{equation}\label{Bom}
B_\Omega^{} \equiv \frac{2 m_e c}{e} \Omega \, \approx \, 1.14 \times 10^{- 7} \left( \frac{\Omega}{\SI{}{Hz} }\right)
\SI{}{G} \, ,
\end{equation}
makes it practically impossible to observe a rotational analog of the Aharonov--Bohm oscillations by varying the rotation frequency, since $\Phi_\Omega/\Phi_0 \ll 1$. 

Existing Aharonov--Bohm interferometers have radii on the order of $R \sim \SI{0,5}{\micro\metre}$ \cite{Russo2008PRB, Ensslin2022NanoLett}, 
whereas experiment  with electrons in vacuum have used enclosed areas of approximately $\pi R^2 \sim \SI{3.9}{mm^2}$  \cite{Hasselbach1993PRA} 
making it approximately $5 \times 10^6$ times more sensitive. 
This is why, to enhance the Sagnac effect, it has been proposed to use a series of $10^6$ to $10^7$ rings in order to achieve a 
signal-to-noise ratio greater than 1 for sub-Hertz rotations \cite{Zivkovic2008PRB, Toland2010PLA, Search2011patent}.
It is worth noting that arrays of rings can also be realized experimentally  \cite{Bleszynski-Jayich2009-Sci}.

The ratio of the Mashhoon to Sagnac fringe shifts can be estimated as
\begin{equation}
\frac{\Theta_\text{M}}{\Theta_\text{S}} = \frac{g_\Omega}{4 \pi}\, \frac{c}{v}\, \frac{\lambda_\mathrm{C} }{R} \approx  1.9 \times 10^{-7}\, 
\frac{c}{v}\, \frac{\SI{}{\micro\metre}}{R} \, . 
\end{equation}
Taking into account that $c/v \approx 300$ for graphene, and using the same value $R = \SI{0.5}{\micro m}$, we estimate the ratio $\Theta_\text{M}/ \Theta_\text{S} \approx 10^{-4}$, which is six orders of magnitude larger than the value reported in Ref.~\cite{Danner2020QI}. This ratio can be further increased by using materials with a low Fermi velocity. For example, a 3D topological insulator Bi$_2$Te$_3$ is a promising candidate due to its notably low Fermi velocity \cite{Wolos2012PRL},
$v \approx \SI{3260}{m/s}$, which is over 300 times smaller than that of graphene. As a result, the ratio $\Theta_\text{M}/ \Theta_\text{S}$ is estimated to be approximately $0.03$.

This raises the question of whether the Mashhoon effect can also be realized in solid-state systems. As discussed above, this requires that electron spins be oriented in opposite directions for waves propagating clockwise and counterclockwise~\cite{Mashhoon1988PRL}. We propose that this configuration can be realized by covering the two arms of the Mach--Zehnder interferometer with ferromagnetic layers having opposite magnetization directions. Techniques for controlling spin orientation in this way are commonly employed in the fabrication of spintronic devices (see, e.g., Refs.~\cite{Han2014-NatNano, Xie2025-mg}).

\section*{Acknowledgements}
We are grateful to G.~Vignale for useful communication and to E.\,V.~Gorbar, V.\,P.~Gusynin, A.\,A.~Kordyuk, V.\,P.~Kravchuk, A.\,A.~Semenov and Y.\,O.~Zolotaryuk for numerous stimulating discussions.


\paragraph{Funding information}
Y.\,V.\,S.\@ is supported by the National Academy of Sciences of Ukraine under project 0126U000353 and by a grant from Simons Foundation International SFI-PD-Ukraine-00014573, PI~LB\@.  T.-H.\,O.\,P.\@ and S.\,G.\,S.\@ acknowledge support from the National Research Foundation of Ukraine grant (2023.03/0097) ``Electronic and transport properties of Dirac materials and Josephson junctions.''

\begin{appendix}
\numberwithin{equation}{section}

\section{Dirac and Pauli equations in a rotating frame}
\label{app:Dirac}

The Dirac equation for a bispinor $\Psi$ describing an electron with vacuum mass $m_e$ and electric charge $- e$ in a gravitational field (or in arbitrary coordinates in flat space-time) reads
\begin{equation} \label{Dirac}
\gamma^a e_a^\mu \left[ {\rm i} \hbar \left( \partial_\mu + \omega_\mu \right) + \frac{e}{c} A_\mu \right] \Psi - m_e c \Psi = 0 \, ,
\end{equation}
where $e_a^\mu$ is the tetrad basis, $A_\mu$ is the covector potential of the external electromagnetic field, and
\begin{equation}
\omega_\mu = \frac18 \omega^{ab}_\mu \left[ \gamma_a , \gamma_b \right] 
\end{equation}
is the spin connection. The Latin indices $a = 0,1,2,3$ refer to the tangent space, and the Greek indices $\mu = 0,1,2,3$ refer to the world coordinates. The matrices $\gamma^a$ are the usual constant Dirac gamma-matrices.

We denote the laboratory inertial coordinates by $\left( c t', x', y', z' \right)$, and the coordinates of the system rotating about the $z'$ axis with constant angular velocity $\Omega$ by the same letters without primes. They are related by the transformation
\begin{equation}
x' = x \cos \Omega t - y \sin \Omega t \, , \qquad y' = y \cos \Omega t + x \sin \Omega t \, , \qquad z' = z \, , \qquad t' = t \, .
\end{equation}
The space-time metric in these coordinates reads
\begin{equation}
d s^2 = c^2 d t'{}^2 - d x'{}^2 - d y'{}^2 - d z'{}^2 = c^2 d t^2 - \left( d x - \Omega y d t \right)^2 - \left( d y + \Omega x d t \right)^2  - d z^2 \, .
\end{equation}

First of all, we need to choose the tetrad basis to which we refer the spinor. In a conventional approach, the tetrad basis is chosen as that of the laboratory frame but rotating about the $z'$ direction with angular velocity $\Omega$, so that the two basis vectors in the rotation plane are directed along the (rotating) $x$ and $y$ axes. For the tetrad $e_a \equiv e_a^\mu \partial_\mu$ expressed in the coordinates of the rotating frame, this gives (this tetrad was also adopted in \cite{Matsuo2011PRL, Chen:2015hfc, Fukushima2019})
\begin{equation} \label{tetrad-v}
c e_0 = \partial_t + \Omega \left( y \partial_x - x \partial_y \right) \, , \qquad e_1 = \partial_x \, , \qquad e_2 = \partial_y\, , \qquad e_3 = \partial_z \, .
\end{equation} 
The spin connection in this tetrad frame can be readily calculated (e.g., by using Cartan's structure equations). The only non-zero component of the spin connection form $\omega^{ab} \equiv \omega^{ab}_\mu d x^\mu$ is
\begin{equation}
\omega^{12} = - \omega^{21} = \Omega\, d t \, .
\end{equation}

The covector potential $A_\mu$ in \eqref{Dirac} is a superposition of the electromagnetic field induced by the lattice and the external field. We assume the presence of an external homogeneous magnetic field with component $B$ along the $z'$-axis in the non-rotating frame. The corresponding electromagnetic field $A^\text{ext} \equiv A^\text{ext}_\mu d x^\mu$ is then 
\begin{equation}
A^\text{ext} = \frac12 B \left( y' d x' - x' d y' \right) = \frac12 B \left[ y d x - x d y - \Omega \left( x^2 + y^2 \right) d t \right] \, .
\end{equation}
Note that $A^\text{ext}_\mu e_0^\mu = 0$.

In the rotating frame, both the metric and the crystal are stationary, so the electromagnetic potential $A_\mu$ can also be chosen to be stationary. For the crystal electric field, we set $A_0 = V$ independent of time, and we neglect other components of the intrinsic covector potential in the rotating frame.  The crystalline structure is rotating as a whole, and we assume that its intrinsic form remains unchanged (in other words, we neglect its intrinsic deformation due to local acceleration). 

Combining all our previous expressions, we can write equation \eqref{Dirac} in the rotating frame in the form
\begin{equation} \label{Dirac-rot}
\left[ \gamma^0 \left( {\rm i} \hbar \partial_t + \Omega L_z + \frac{{\rm i} \hbar}{4} \Omega \left[ \gamma_1, \gamma_2 \right] + e V \right)  + c \gamma^i \left( {\rm i} \hbar \partial_i + \frac{e}{c} A_i \right) \right] \Psi - m_e c^2 \Psi = 0 \, ,
\end{equation}
where $L_z = {\rm i} \hbar \left( y \partial_x - x \partial_y \right)$ is the operator of the $z$-component of angular momentum, and
\begin{equation}\label{Acomp}
A_i dx^i = \frac12 B \left( y d x - x d y \right)
\end{equation} 
is the covector potential of the external magnetic field as it appears in the rotating frame.

We are interested in the non-relativistic limit of Eq.~\eqref{Dirac-rot}. This is most conveniently done in the so-called standard representation of the Dirac gamma-matrices \cite[\S\,21]{Berestetskii:1982qgu}:
\begin{equation}
\beta \equiv \gamma^0 = \begin{pmatrix} 1 & 0 \\ 0 & -1 \end{pmatrix} \, , \qquad \gamma^i = \begin{pmatrix} 0 & \sigma_i \\ - \sigma_i & 0 \end{pmatrix} \, , \qquad \alpha^i \equiv \gamma^0 \gamma^i = \begin{pmatrix} 0 & \sigma_i \\ \sigma_i & 0 \end{pmatrix} \, , 
\end{equation}
where $\sigma_i$ are the two-dimensional Pauli matrices. We have
\begin{equation}
\left[ \gamma_i , \gamma_j \right] = - 2 {\rm i}_{} \epsilon_{ijk} \Sigma_k \, , \qquad \Sigma_k = \begin{pmatrix} \sigma_k & 0 \\ 0 & \sigma_k \end{pmatrix} \, . 
\end{equation}

According to the standard derivation \cite[\S\,33]{Berestetskii:1982qgu} of the leading non-relativistic limit, we introduce the two-component spinors $\psi$ and $\chi$ such that
\begin{equation}
\Psi = e^{- {\rm i}_{} m_e c^2 t / \hbar} \begin{pmatrix} \psi \\ \chi \end{pmatrix} \, .
\end{equation}
Then, in the leading non-relativistic approximation, the spinor $\psi$ satisfies the Pauli equation in the rotating frame:
\begin{equation} \label{Pauli}
\left( {\rm i} \hbar \partial_t + \Omega J_z \right) \psi = \left[ \frac{1}{2 m_e} \sum_i \left( {\rm i} \hbar \partial_i + \frac{e}{c} A_i \right)^2 - e V + \frac{e \hbar}{2 m_e c} B \sigma_z \right] \psi \, ,
\end{equation}
with the operator of total angular momentum $J_z = L_z + \hbar \sigma_z / 2$, and $\sigma_z = \sigma_3$.

\section{Equivalence of rotation and magnetic field}
\label{app:Larmor}

The Pauli equation \eqref{Pauli} can be written in the form
\begin{equation}\label{eff}
{\rm i} \hbar \partial_t \psi = \left[ - \frac{\hbar^2}{2 m_e} \vec \nabla{}^2 - e V - \left( \Omega - \frac{e}{2 m_e c} B \right) L_z + \frac{e^2 B^2 r^2}{8 m_e c^2} - \left( \Omega - \frac{e}{2 m_e c} B \right) \hbar \sigma_z + \frac{\hbar}{2} \Omega \sigma_z \right] \psi \, ,
\end{equation}
where $r^2 = x^2 + y^2$. 

Introducing the effective field
\begin{equation}\label{Beff}
B_\text{eff} = B - \frac{2 m_e c}{e} \Omega \, ,
\end{equation}
we write Eq.~\eqref{eff} as
\begin{equation} \label{eff-fin}
{\rm i} \hbar \partial_t \psi = \left[ H \left( B_\text{eff} \right) - \frac12 m_e \Omega^2 r^2 + \frac{e}{2 c} B \Omega r^2 + \frac{\hbar}{2} \Omega \sigma_z \right] \psi \, ,
\end{equation}
where $H \left( B \right)$ denotes the electronic Hamiltonian in the material at rest in a magnetic field $B$:
\begin{equation}\label{Hom}
H \left( B \right) \equiv - \frac{\hbar^2}{2 m_e} \vec \nabla{}^2 - e V + \frac{e^2 B^2 r^2}{8 m_e c^2} + \frac{e}{2 m_e c} B \left( L_z + \hbar \sigma_z \right)\, .
\end{equation}

Equation \eqref{eff-fin} includes both the magnetic field--rotation and spin--rotation couplings, represented by the last two terms, respectively, which also appear in the non-relativistic treatment presented in \cite{Vignale1995PL}. In the absence of an external magnetic field ($B = 0$), the operator on the right-hand side of \eqref{eff-fin} differs from the Hamiltonian $H \left( B_\text{eff} \right)$\,---\,defined in \eqref{Hom} and describing the material at rest in the effective external magnetic field $B_\text{eff}$\,---\,only by the presence of an additional centrifugal potential energy term $- m_e \Omega^2 r^2 / 2$, which can be combined with the electric potential energy $- e V$, and by the spin--rotation coupling term $\hbar \Omega \sigma_z / 2$. These minor differences between the two Hamiltonians represent another form of the well-known Larmor equivalence theorem \cite{Vignale1995PL}. It should be noted that the effective mass $m^\ast$ emerges as a result of considering the electron motion within the lattice potential $V$.

By virtue of this Larmor theorem, the Sagnac effect for any system can be interpreted as a spin-dependent Aharonov--Bohm effect in the presence of the effective magnetic field given by \eqref{Beff}, together with the extra terms in \eqref{eff-fin}. Notably, it is the vacuum electron mass that appears in \eqref{Beff}. Similar conclusions were drawn in \cite{Vignale1995PL} by analyzing the classical form of a non-relativistic electronic Hamiltonian in a rotating frame.

In general, equations \eqref{eff} or \eqref{eff-fin} show that, to linear order in the field or rotation frequency, the Larmor correspondence \eqref{Beff} holds for orbital motion. However, for spin, the correspondence between the magnetic field and rotation has a different form due to the last term in \eqref{eff} or \eqref{eff-fin}. This justifies our approach in Sec.~\ref{sec:nonrel}, in which we apply the Larmor theorem to the orbital motion (extending spatial partial derivatives by the corresponding effective vector potential), while introducing the $g$-factor for intrinsic spin.

We recall that the $g$-factor $g_B$ associated with the magnetic field appears in the effective spin Hamiltonian, defined as
\begin{equation}
H_{B\,\text{eff}} = \mu_\text{B} g_B B S_z / \hbar \, ,
\end{equation}
where $\mu_\text{B} = e \hbar / 2 m_e c$ is the Bohr magneton, and $S_z = \hbar \sigma_z / 2$ is the spin operator (see \cite{Ihn2010book, Prada:2021} and references therein). It is measured in electron spin resonance experiments and should meet the resonance condition
\begin{equation}\label{g-mag}
\left\langle L_z + g_0 S_z \right\rangle = g_B \left\langle S_z \right\rangle \, ,
\end{equation}  
where $g_0 = 2$ is the magnetic $g$-factor for a free electron in Dirac's theory, and the angle brackets indicate the expectation value with respect to the electronic state in the solid.

In quite a similar way, we introduce a rotational $g$-factor $g_\Omega$ in the effective Hamiltonian describing the coupling between rotation and spin:
\begin{equation} \label{g-Omega}
H_{\Omega\, \text{eff}} = - g_\Omega \Omega S_z \, .
\end{equation}
Examining the $\Omega$-dependent coupling in Eq.~\eqref{eff}, we observe that it will be given by a relation similar to \eqref{g-mag}:
\begin{equation} \label{g-rot}
\left\langle L_z + g_0 S_z - S_z \right\rangle = g_\Omega \left\langle S_z \right\rangle \, .
\end{equation}

Combining Eqs.~\eqref{g-mag} and \eqref{g-rot}, we obtain a noteworthy relation between the magnetic and rotational $g$-factors:
\begin{equation}\label{g-1}
g_\Omega = g_B - 1 \, .
\end{equation}
For a free electron, this relation is naturally satisfied: $g_B = 2$ while $g_\Omega = 1$.
\end{appendix}





\bibliography{sagnac-new.bib}


\end{document}